\documentclass{aa}
\usepackage{graphicx}
\usepackage{txfonts}

\begin{document}

   \title{The X-ray/radio and UV luminosity expected from  symbiotic systems as the progenitor of SNe Ia}


   \author{Xiangcun Meng
          \inst{1,2}
          and
          Zhanwen Han \inst{1,2}}

   \offprints{X. Meng \& Z. Han}

   \institute{$^{\rm 1}$Yunnan
Observatories, Chinese Academy of Sciences, Kunming, 650216,
China\\
              \email{[xiangcunmeng, zhanwenhan]@ynao.ac.cn}\\
$^{\rm 2}$Key Laboratory for the Structure and Evolution of
Celestial
Objects, Chinese Academy of Sciences, Kunming 650216, China\\
             }

   \date{Received; accepted}


  \abstract
   {Symbiotic systems (i.e. a white dwarf + red giant star, WD + RG),
   which experience mass loss and form circumstellar material
   (CSM), have been suggested as being a possible progenitor system
   of type Ia supernovae (SNe Ia). After a supernova explosion, the
   supernova ejecta may interact with the CSM or the RG secondary.
   X-ray/radio emission (excess
   UV photons) is expected from the interaction between supernova ejecta and the CSM (RG secondary).
   However, no X-ray or radio emission that has originated from this type of system has been observationally detected, and
   only four SNe Ia have shown any possible signal of excess UV
   emission. These observational discrepancies need to be interpreted.
   }
   {We seek to determine the luminosity of these emissions, using detailed
   binary evolution algorithms to obtain the parameters of binary systems at the moment of the supernova explosion.
}
   {We carried out a series of binary stellar evolution calculations, in
which the effect of tidally enhanced wind on the evolution of WD +
RG systems is incorporated. The WDs increase their mass to the
Chandrasekhar mass limit, and then explode as SNe Ia. Based on the
binary evolution results, we estimated the X-ray/radio (the excess
UV) luminosity from the interactions between supernova ejecta and
the CSM (the secondary) using a variety of  published standard
models.}
   {We found that the X-ray flux may be high enough to be detected for a nearby SN Ia from a symbiotic system,
   while the radio flux is more likely to de detected when
   the companion is an asymptotic giant branch (AGB) star, and for a first giant branch (FGB) companion,
   the radio flux is generally lower than the detection limit. For two well observed SNe Ia,
   2011fe and 2014J, almost all symbiotic systems
   are excluded by X-ray observations, but WD + FGB systems may not be ruled out by radio observations. The excess UV luminosity
   that results from the collision of supernova ejecta with the RG secondary may be high enough
   to be detected if the secondary fills its Roche lobe at the moment of a supernova explosion.}
   {The X-ray/radio emissions are more prevalent in SNe Ia from WD + AGB
   systems, although SNe Ia from such systems are rare.
   The UV luminosity from the collision of supernova ejecta to RG secondary is high
   enough, but only one in every few hundred SNe Ia manifests the signal from the collision.}

   \keywords{binaries: symbiotic - stars: evolution - mass-loss -
supernovae: general - white dwarfs
               }
   \authorrunning{Meng \& Han}
   \titlerunning{The X-ray/radio and UV luminosity expected from WD + RG systems}
   \maketitle{}
%

\section{Introduction}\label{sect:1}
Type Ia supernovae (SNe Ia) play an important role in many
astrophysical fields, especially in cosmology. SNe Ia appear to be
good cosmological distance indicators and have been successfully
applied in determining cosmological parameters (e.g., $\Omega$ and
$\Lambda$), which resulted in the discovery of the accelerating
expansion of the universe (Riess et al. \cite{RIE98};  Schmidt et
al. \cite{SCHMIDT98}; Perlmutter et al. \cite{PER99}). SNe Ia are
also  proposed as  cosmological probes to test the evolution of
the dark energy equation of state with time (Howell et al.
\cite{HOWEL09}; Meng et al. \cite{MENG15}). However, the
progenitor systems of SNe Ia have not yet been confidently
identified (Hillebrandt \& Niemeyer \cite{HN00}; Leibundgut
\cite{LEI00}; Wang \& Han \cite{WANGB12}; Maoz, Mannucci \&
Nelemans \cite{MAOZ14}), although the identification of the
progenitor is crucial in many astrophysical fields, e.g., in
determining cosmological parameters, studying galaxy chemical
evolution, understanding the explosion mechanism of an SN Ia, and
constraining the theory of binary stellar evolution (see the
review by Livio \cite{LIVIO99})

It is widely believed that an SN Ia results from the explosion of
a carbon--oxygen white dwarf (CO WD) in a binary system (Hoyle \&
Fowler \cite{HF60}). The CO WD may increase its mass by accreting
from its companion to a maximum stable mass, and then explode as a
thermonuclear runaway. A large amount of radioactive nickel-56
($^{\rm 56}$Ni) is produced in the explosion (Branch
\cite{BRA04}), and the amount of $^{\rm 56}$Ni determines the
maximum luminosity of the SN Ia (Arnett \cite{ARN82}). Based on
the nature of the companion of the mass accreting WD, two basic
progenitor models for SN Ia have been discussed over the last
three decades. One is the single degenerate (SD) model (Whelan \&
Iben \cite{WI73}; Nomoto, Thielemann \& Yokoi \cite{NTY84}), in
which a CO WD accretes hydrogen- or helium-rich materials from its
companion to increase its mass, and finally explodes if its mass
approaches the Chandrasekhar mass limit. The companion may be
either a main-sequence (MS), a slightly evolved star (WD+MS) or a
red giant star (WD+RG), or a helium  star (WD + He star; Yungelson
et al. \cite{YUN95}; Li \& van den Heuvel \cite{LI97}; Hachisu et
al. \cite{HAC99a,HAC99b}; Nomoto et al. \cite{NOM99, NOM03};
Langer et al. \cite{LAN00}; Han \& Podsiadlowski \cite{HAN04,
HAN06}; Chen \& Li \cite{CHENWC07, CHENWC09}; Meng, Chen \& Han
\cite{MENG09}; L\"{u} et al. \cite{LGL09}; Wang et al.
\cite{WANGB09a, WANGB09b}). An alternative is the double
degenerate (DD) model (Iben \& Tutukov \cite{IT84}; Webbink
\cite{WEB84}; Han \cite{HAN98}; Chen et al. \cite{CHENXF12}), in
which a system that consists of two CO WDs loses orbital angular
momentum by gravitational wave radiation and ultimately merges.
The merger may explode if the total mass of the system exceeds the
Chandrasekhar mass limit (see the reviews by Hillebrandt \&
Niemeyer \cite{HN00} and Maoz, Mannucci \& Nelemans
\cite{MAOZ14}). Apart from these two basic scenarios, several
other models have also been frequently discussed by many groups,
such as the core-degenerate model (Kashi \& Soker \cite{KASHI11};
Ilkov \& Soker \cite{ILKOV12}), the WD$-$WD collision model
(Raskin et al. \cite{RASKIN09}; Kushnir et al. \cite{KUSHNIR13})
and the double-detonation scenarios (Woosley \& Weaver
\cite{WOOSLEY94}, Livne \& Arnett \cite{LIVIN95}; Ruiter et al.
\cite{RUITER11}), and one may refer to Tsebrenko \& Soker
(\cite{TSEBRENKO15}) for further information. At present, it is
premature to exclude any of these channels.

The symbiotic channel (WD + RG), where the CO WD mainly increases
its mass by wind accretion, is believed to be a possible channel
for producing SNe Ia (Branch et al. \cite{BRANCH95}; Hachisu et
al. \cite{HAC99b}), but some observations seem to disfavour this
channel, e.g., no radio emission, which is expected from the
interaction between supernova ejecta and lost material from the
binary system, was detected, although   other
observations  exist  that uphold the channel. (Please see Section
\ref{sect:2} for observations.) In this paper, we  try to
interpret  observations that disfavor the WD + RG channel,
and demonstrate that these observational results are expected to
occur naturally as a consequence of this channel.

In Sect. \ref{sect:2}, we summarize various observational facts. In
Sect. \ref{sect:3}, we describe our model and present the
calculation results in Sect. \ref{sect:4}. We provide discussions in
Sect. \ref{sect:5} and draw our conclusions in Sect. \ref{sect:6}.


\section{Observations and models}\label{sect:2}
\subsection{Observations}\label{sect:2.1}
There are at least four recurrent symbiotic novae, such as RS Oph
and T CrB, whose WD masses are very massive (Mikolajewska
\cite{MIKOLAJEWSKA10,MIKOLAJEWSKA11}). The massive WDs may
increase their mass by wind accretion, from which SNe Ia can
arise (Hachisu \& Kato \cite{HK01}; Sokoloski et al.
\cite{SOKOLOSKI06a}; Podsiadlowski \cite{PODSIADLOWSKI08}). If
some symbiotic novae are reliable progenitor systems for SNe Ia,
we can expect certain special observational properties in SNe Ia
to be as follows:

1) Prior to the supernova explosion, the systems must lose a large
amount of material to form circumsteller material (CSM). Although
the expected UV flux is, in general, expected to be low, there may
be enough flux from the photosphere to ionize atoms in the CSM
with low ionization potential. If the SN ejecta were to interact
with the CSM, or impact an RG, the amount of UV flux may be
disputable (Kasen \cite{KASEN10}; Lundqvist et al.
\cite{LUNDQVIST13}), and a large fraction of the CSM will be
ionized. As the number of UV photons  varies with time, we can
expect to see variable circumstellar absorption lines in the
spectra of SNe Ia. Variable circumstellar Na I absorption lines
were first observed in the spectra of SN 2006X (Patat et al.
\cite{PAT07}). The distance from the explosion center to the
absorbing material was estimated as being less than $10^{\rm 17}$
cm (Patat et al. \cite{PAT07}; Simon et al. \cite{SIMON09}). The
variable circumstellar absorption lines have also been observed in
RS Oph (Patat et al. \cite{PATAT11}), which linked the symbiotic
system to SNe Ia. Judging by the expansion velocity of the CSM,
Patat et al. (\cite{PAT07}) suggested that the progenitor of SN
2006X is a WD + RG system, although the possibility of a WD + MS
system cannot be excluded completely (Hachisu et al.
\cite{HKN08}). Subsequently, two twins of SN 2006X have also been
reported (Blondin et al. \cite{BLONDIN09}; Simon et al.
\cite{SIMON09}). Recently, Graham et al. (\cite{GRAHAM15}) have
found that several spectra with variable K I absorption lines and
they argue that the most plausible explanation for this is the
photoionization of CSM, but the origin of the CSM is controversial
(Soker \cite{SOKER15}; Maeda et al. \cite{MAEDA16}). In addition,
owing to the existence of the CSM, the possibility of significant
blueshifts for some absorption lines can be expected in the
spectra of some SNe Ia, even though the true nature of the
blueshift is not fully understood (Sternberg et al.
\cite{STERNBERG14}). Sternberg et al. (\cite{STERNBERG11}) checked
the absorption properties of 35 SNe Ia and found that the velocity
structure of absorbing material along the line of sight for more
than 50\% of  Ia supernovae tends to be blueshifted. These
structures are likely to be the signatures of gas outflows from
the SD systems, although it is possible for some DD systems to
show  signs of CSM as well, e.g., Shen et al. (\cite{SHEN13}).
Furthermore, Foley et al. (\cite{FOLEY12}) even find that SNe Ia
with blueshifted circumstellar/interstellar absorption
systematically have higher ejecta velocities, which could mean
that the SNe Ia comes from a variety of progenitor systems and
that  SNe Ia from systems with possible winds tends to have more
kinetic energy per unit mass than those from  systems with weak or
no winds. Maguire et al. (\cite{MAGUIRE13}) also noticed that  SNe
Ia that display blue-shifted absorption features are skewed
towards later-type galaxies and have broader light curves,
compared to those without absorption features.

2) The wind from  symbiotic systems may interact with interstellar
material and this kind of interaction may reveal itself during the
remnant stage. Chiotellis et al. (\cite{CHIOTELLIS12}) present a
model about the remnant of  Kepler's supernova and find that the
main features of the remnant can be explained by a symbiotic
binary, which consists of a white dwarf and an asymptotic giant
branch (AGB) donor star with an initial mass of 4-5 $M_{\odot}$.
In their model, the interaction between the stellar wind and the
interstellar material plays a very important role in explaining
the X-ray properties of Kepler's supernova remnant (see also
Patnaude et al. \cite{PATNAUDE12} and Burkey et al.
\cite{BURKEY13}). Recently, on the basis of a \emph{Spitzer}
24$\mu$ image, Williams et al. (\cite{WILLIAMS14}) have linked the
SNR 0509-68.7 in the Large Magellanic Cloud (LMC) to pre-supernova
mass loss from a possible SD system, where the companion may have
beeen a relatively unevolved star at the time of the explosion
owing to a lack of N-enhancement.

3) After the supernova explosion, the explosion ejecta runs into
the CSMs and interacts with them. The interacting materials may
emit at radio band and X-ray (Chevalier \cite{CHEVALIER90}).
Therefore, radio and X-ray observation may help to discriminate
between different progenitor models by shedding light on the
properties of the environment that was shaped by the evolution of
the progenitor system since these forms of radiation are generally
not expected from the DD model (Boffi \& Branch \cite{BOFFI95};
Eck et al. \cite{ECK95}). However, see Shen et al. (\cite{SHEN13})
for a different view. Motivated by this, several dozen SNe Ia have
been observed in radio (Panagia et al. \cite{PANAGIA06}; Hancock
et al. \cite{HANCOCK11}) and in X-rays (Hughes et al.
\cite{HUGHES07}; Immler et al. \cite{IMMLER06}; Russell \& Immler
\cite{RUSSELL12}; Margutti et al. \cite{MARGUTTI12}). No SN Ia has
been ever detected in radio and in X-rays, except perhaps for a
possible detection of SN 2005ke (Immler et al. \cite{IMMLER06},
which was disputed by Hughes et al. \cite{HUGHES07}), and which
seems to exclude evolved stars as the companion objects in the
progenitor systems of SNe Ia, such as symbiotic systems. The
progenitor model of an SNe Ia should try to explain these
observations, especially for the wind accretion model from WD + RG
systems.

4)After the supernova explosion, the supernova ejecta may collide
into the RG companion whereby the interaction may produce an
excess of UV radiation (Kasen \cite{KASEN10}). This kind of UV
excess was expected in early UV observations. In any case,
reporting the detection of  UV excess is usually negative (Hayden
et al. \cite{HAYDEN10}; Tucker \cite{TUCKER11}; Bianco et al.
\cite{BIANCO11}; Brown et al. \cite{BROWN12b}; Olling et al.
\cite{OLLING15}), which seems to exclude a red supergiant
companion, at least in the case of a Roche-lobe overflow. However,
Ganeshalingam et al. (\cite{GANESHALINGAM11}) find that there is a
substantial degree of degeneracy between the adopted power-law
index of the SN light-curve template, the rise time, and the
amount of shock emission required to match the data, and that a
conclusive result is difficult to obtain. Recently, there have
been four positive reports for this kind of UV excess detection,
which supports the WD + RG systems (2009ig, Foley et al.
\cite{FOLEY12b}; 2011de, Brown \cite{BROWN14}; iPTF14atg, Cao et
al. \cite{CAOY15}) and systems with massive MS companions (SN
2012cg, Marion et al. \cite{MARION15}) as possible progenitor
systems. In addition, the early tail of the light curves of 35
high-z SNe Ia, which was  found by the Supernova Cosmology Project
in Goldhaber et al. (\cite{GOLGHABER01}) definitely looks
different from the lack-of-companion prediction, although the RG
companions does not seem to be favored by these data (see the
review by Ruiz-Lapuente \cite{RUIZLAPUENTE14}). However, these
observations raise a question, i.e. why do  so few SNe Ia
definitely show signs of  UV excess  among the hundreds of SNe Ia?
In addition to the direct impact, surely evaporated material from
the RG/MS star would be heated and excited by the radioactive
input from the supernova, and  show up after about one year in the
form of emission lines with a width of $\sim10^{\rm 3}$ km/s
(Marietta et al. \cite{MAR00}; Meng et al. \cite{MENGXC07}; Pan et
al. \cite{PAN12}). So far, no such lines have been detected for
the seven SNe Ia discussed in the literature (Mattila et al.
\cite{MATTILA05}; Leonard \cite{LEONARDO07}; Lundqvist et al.
\cite{LUNDQVIST13, LUNDQVIST15}). Based on modeling, none of these
SNe seem to have been the result of a WD + RG system (see
Lundqvist et al. \cite{LUNDQVIST15}). This agrees with few cases
of observed UV excess among SNe Ia.

In the above observations, Observations (1) and (2) favor WD + RG, while
observation (3) disfavors them. Observation (4) brings out
an interesting question which needs to be addressed by the SD
model. We  show in theory that actually, observations (3) and
(4) are natural results in both physics and statistics.

\subsection{Previous binary evolution model}\label{sect:2.2}
Although there are some symbiotic systems with large massive WDs,
it is difficult for these systems to produce SNe Ia by standard
binary evolution because the mass transfer would be dynamically
unstable when the RG companions fill their Roche lobe. To overcome
this problem, some mechanisms were suggested. By a simple
analytical method to treat binary interactions, Hachisu et al.
(\cite{HAC99b}) studied the binary evolution of WD + MS and WD +
RG systems. In their study, a mass-stripping effect from an
optically thick wind (OTW) was introduced. The effect may
attenuate the mass-transfer rate from the companion to the WD, and
was adopted by way of explaining some quasi-regular SSSs (Hachisu \& Kato
\cite{HK03a,HK03b}). According to these studies, Hachisu
(\cite{HKN08b}) produced a power-law DTD that follows on the equation
(1) of Iben \& Tutukov (\cite{IT84}), i.e.

 \begin{equation}
 \nu=0.2\Delta q\int_{M_{\rm A}}^{M_{\rm B}}\frac{dM}{M^{\rm 2.5}}\Delta\log A
 \hspace{0.2cm}{\rm yr^{\rm -1}},\label{eq:IT84}
  \end{equation}
where $\Delta q$, $\Delta\log A$, $M_{\rm A}$, and $M_{\rm B}$ are
the appropriate ranges of the initial mass ratio and separation,
and the lower and upper limits of the primary mass for producing
SNe Ia in units of solar masses, respectively. However, the birth
rate from this equation is probably overestimated since some
parameter spaces, which are considered as producing SNe Ia from this equation,
may not contribute to SNe Ia. Instead of the mass-stripping
effect, Chen et al. (\cite{CHENXF11}) constructed a
tidally-enhanced wind (TEW) model, which may effectively avoid the
dynamically instable mass transfer between the WD and its RG
companion, and produce SNe Ia with a very long delay time. In this
paper, our method for binary evolution is similar to that in Chen
et al. (\cite{CHENXF11}) (see also Meng \& Podsiadlowski
\cite{MENGPOD13}).

\section{Model and method}\label{sect:3}

\subsection{Physical Input}\label{sect:3.1}
 We use the stellar evolution code of Eggleton
(\cite{EGG71, EGG72, EGG73}) to calculate the binary evolutions of
SD systems. The code has been updated with the latest input
physics from over the last four decades (Han, Podsiadlowski \&
Eggleton \cite{HAN94}; Pols et al. \cite{POL95, POL98}).
Roche-lobe overflow (RLOF) is treated within the code described by
Han et al. (\cite{HAN00}). We set the ratio of mixing length to
local pressure scale height, $\alpha=l/H_{\rm p}$, to 2.0, and set
the convective overshooting parameter, $\delta_{\rm OV}$, to 0.12
(Pols et al. \cite{POL97}; Schr\"{o}der et al. \cite{SCH97}),
which roughly corresponds to an overshooting length of $0.25
H_{\rm P}$. The solar metallicity is adopted here ($Z=0.02$). The
opacity table for the metallicity is compiled by Chen \& Tout
(\cite{CHE07}) from Iglesias \& Rogers (\cite{IR96}) and Alexander
\& Ferguson (\cite{AF94}).

\subsection{Wind accretion}\label{sect:3.2}
If the WD companion is a main-sequence (MS) star, the velocity of
the outflow from the binary system is usually very high, in which case
it would be very difficult to detect the CSM. Therefore, we only
consider the case of the red-giant (RG) companion in this paper.
The method that calculates binary evolution here is similar to that in
Chen et al. (\cite{CHENXF11}) and Meng \& Podsiadlowski
(\cite{MENGPOD13})

We assume that the stellar wind mass-loss rate of a secondary in a
binary system is increased by the presence of the WD companion
star, i.e. the tidal enhancement of the mass-loss rate from the
secondary which is modelled by Reimers' (\cite{REIMERS75}) formula
with the extra tidal term included by Tout \& Eggleton
(\cite{TOUT88}) so that

 \begin{equation}
 \begin{array}{lc}
 \dot{M}_{\rm 2w}=-4\times 10^{\rm -13}\frac{\eta(L/L_{\odot})(R/R_{\odot})}{(M_{\rm 2}/M_{\odot})}\\
\hspace{0.75cm}\times\left[1+B_{\rm
W}\min\left(\frac{1}{2},\frac{R}{R_{\rm L}}\right)^{\rm 6}\right]
M_{\odot}{\rm yr}^{\rm -1},
\end{array}
  \end{equation}
where $L$ and $R$ are the luminosity and radius of the giant
secondary, $R_{\rm L}$ is its Roche-lobe radius, and $\eta$ is
 the Reimers¡¯ wind coefficient, which is set at $0.25$. The
wind enhancement factor $B_{\rm W}$ is still uncertain, where it
is more than 3 000 in Tout \& Eggleton (\cite{TOUT88}), and is 10
000 in the wind-driven mass transfer theory of Tout \& Hall
(\cite{TOUT91}). Chen et al. (\cite{CHENXF11}) have shown that the
parameter space leading to SNe Ia increases with $B_{\rm W}$, i.e.
the birth rate of SNe Ia from the WD + RG channel increases with
$B_{\rm W}$. However, $B_{\rm W}$ does not affect other properties
of the WD + RG systems that lead to SNe Ia at the moment of
supernova explosion, such as the mass-loss rate (Meng \&
Podsiadlowski \cite{MENGPOD13}). Here, we set $B_{\rm W}=10  000$,
which means that the parameter space for SNe Ia here is an upper
limit, i.e. the results here include all possibilities resulting
from the uncertainties of $B_{\rm W}$, and that the mass-loss rate
from the secondary could be 150 times as large as  the Reimers¡¯
rate, when the star nearly fills its Roche lobe.

Some of the material lost in the form of stellar wind from the
secondary may be accreted by the WD and the mass accretion rate is
expressed by Boffin \& Jorissen (\cite{BOFFIN88}) as

 \begin{equation}
 \dot{M}_{\rm 2a}=-\frac{1}{\sqrt{1-e^{\rm 2}}}\left(\frac{GM_{\rm WD}}{v_{\rm w}^{\rm 2}}\right)^{\rm 2}\frac{\alpha_{\rm acc}\dot{M}_{\rm 2w}}{2a^{2}(1+v_{\rm orb}^{\rm 2}/v_{\rm w}^{\rm
 2})^{3/2}},\label{eq:ac}
  \end{equation}
where $v_{\rm orb}=\sqrt{G(M_{\rm 2}+M_{\rm WD})/a}$ is the
orbital velocity, $G$ is Newton's gravitational constant, $a$ is
the semi-major axis of the orbit, and $e$ is its eccentricity. In
this paper, we take $e=0$. The accretion efficiency $\alpha_{\rm
acc}$ is set to be 1.5. For simplicity, the wind velocity is set
to be a constant (500 ${\rm km}$ ${\rm s^{\rm -1}}$ in MS stage
and $5$ ${\rm km}$ ${\rm s^{\rm -1}}$ in RGB stage). Wind velocity
may significantly affect the accretion rate of the WDs as well as
the parameter space that leads to the SNe Ia or the birth rate of SNe Ia
from the WD + RG channel. Broadly speaking, the parameter space for SNe
Ia sharply decreases with  the wind velocity. Here, $5$ ${\rm km}$
${\rm s^{\rm -1}}$ is a lower limit for the wind velocity, which
means an upper limit of the parameter space that leads to SNe Ia,
i.e. the results here include all possibilities resulting from the
uncertainties of $v_{\rm w}$. Actually, both $B_{\rm W}$ and
$v_{\rm w}$ are poorly known and we will discuss their effects on
the results in Section \ref{subs:5.3}. In  Equation
(\ref{eq:ac}), if $a$ is small enough, the right-hand side could
be larger than $-\dot{M}_{\rm 2w}$, and then we limit that
$\dot{M}_{\rm 2a}\leq-\dot{M}_{\rm 2w}$ as did by Chen et al.
(\cite{CHENXF11}).

Wind accretion is the only way to transfer material from the
RG secondary to the WD before Roche-lobe overflow (RLOF) begins,
in which case the mass-transfer rate is $\dot{M}_{\rm tr}=\dot{M}_{\rm
2a}$. After RLOF occurs, the material is transferred by both a
stream and by wind so that $\dot{M}_{\rm tr}=\dot{M}_{\rm
2a}+|\dot{M}_{\rm 2RLOF}|$, where $\dot{M}_{\rm 2RLOF}$ is the
mass-transfer rate by RLOF.

\subsection{WD mass growth}\label{sect:3.3}
We adopt the prescription of Hachisu et al. (\cite{HAC99a}) and
Hachisu et al. (\cite{HKN08}) on WDs accreting hydrogen-rich
material from their companions (see also Han \& Podsiadlowski
\cite{HAN04} and Meng, Chen \& Han \cite{MENG09}). The following
is a brief introduction of this prescription. If the WD accretion
rate, $\dot{M}_{\rm tr}$, exceeds a critical value, $\dot{M}_{\rm
c}$, we assume that a part of the accreted hydrogen burns steadily
on the surface of the WD and is converted into helium at the rate
of $\dot{M}_{\rm c}$. The unprocessed matter is assumed to be lost
from the system as an optically thick wind at a rate of
$\dot{M}_{\rm wind}=\dot{M}_{\rm tr}-\dot{M}_{\rm c}$ (Hachisu et
al. \cite{HAC96}). The critical accretion rate is

 \begin{equation}
 \dot{M}_{\rm c}=5.3\times 10^{\rm -7}\frac{(1.7-X)}{X}(M_{\rm
 WD}-0.4) M_{\odot} {\rm yr}^{\rm -1},
  \end{equation}
where $X$ is hydrogen mass fraction and $M_{\rm WD}$ is the mass
of the accreting WD (mass is in $M_{\odot}$ and mass-accretion
rate is in $M_{\odot}{\rm yr}^{\rm -1}$, Hachisu et al.
\cite{HAC99a}).

We adopt the following assumptions when $\dot{M}_{\rm tr}$ is
smaller than $\dot{M}_{\rm c}$. (1) When $\dot{M}_{\rm tr}$ is
higher than $\frac{1}{2}\dot{M}_{\rm c}$, the hydrogen-shell
burning is steady and no mass is lost from the system. At this
stage, the system may appear to be a supersoft X-ray source (SSS)
(2) When $\dot{M}_{\rm tr}$ is lower than $\frac{1}{2}\dot{M}_{\rm
c}$ but higher than $\frac{1}{8}\dot{M}_{\rm c}$, a very weak
shell flash is triggered but no mass is lost from the system. At
this stage, the system may be a recurrent nova (RN) (3) When
$\dot{M}_{\rm tr}$ is lower than $\frac{1}{8}\dot{M}_{\rm c}$, the
hydrogen-shell flash is so strong that no material is accumulated
on to the surface of the CO WD. We define the growth rate of the
mass of the helium layer under the hydrogen-burning shell as
 \begin{equation}
 \dot{M}_{\rm He}=\eta _{\rm H}\dot{M}_{\rm tr},
  \end{equation}
where $\eta _{\rm H}$ is the mass-accumulation efficiency for
hydrogen burning. According to the assumptions above, the values
of $\eta _{\rm H}$ are:

 \begin{equation}
\eta _{\rm H}=\left\{
 \begin{array}{ll}
 \dot{M}_{\rm c}/\dot{M}_{\rm tr}, & \dot{M}_{\rm tr}> \dot{M}_{\rm
 c},\\
 1, & \dot{M}_{\rm c}\geq \dot{M}_{\rm tr}\geq\frac{1}{8}\dot{M}_{\rm
 c},\\
 0, & \dot{M}_{\rm tr}< \frac{1}{8}\dot{M}_{\rm c}.
\end{array}\right.
\end{equation}
Therefore, based on the prescription above, the system may
experience one, two, or three of the OTW, SSSs, and RNe stages. In
other words, the supernova may explode in the OTW, SSSs, or RNe
stage.

Helium is ignited when a certain amount of helium is accumulated.
If an He flash occurs, some of the helium is blown off from the
surface of the CO WD. Then, the mass growth rate of the CO WD,
$\dot{M}_{\rm WD}$, is
 \begin{equation}
 \dot{M}_{\rm WD}=\eta_{\rm He}\dot{M}_{\rm He}=\eta_{\rm He}\eta_{\rm
 H}\dot{M}_{\rm tr},
  \end{equation}
where $\eta_{\rm He}$ is the mass-accumulation efficiency for
helium-shell flashes, and its value is taken from Kato \& Hachisu
(\cite{KH04}).

   \begin{figure}
   \centering
   \includegraphics[width=60mm,height=80mm,angle=270.0]{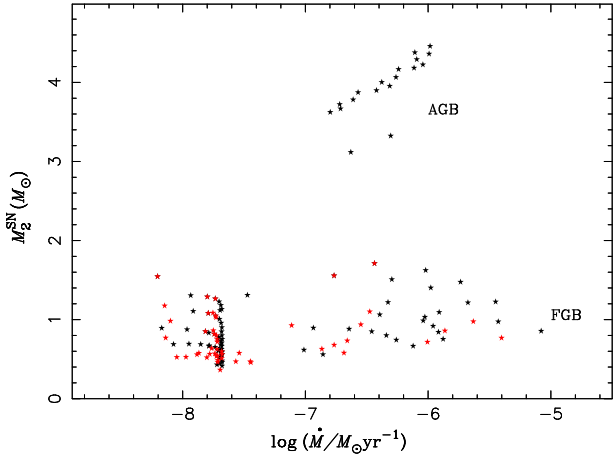}
   \caption{Final companion mass and mass-loss rate from the system at the moment of supernova explosion.
   The red and black points are from the cases of $M_{\rm WD}^{\rm i}=1.0M_{\odot}$ and $M_{\rm WD}^{\rm i}=1.1M_{\odot}$,
   respectively. The companions in the top group are AGB stars, while first giant branch (FGB) stars appear in the bottom group.} \label{m2mdot}%
    \end{figure}

We incorporated all the above prescriptions into Eggleton's
stellar evolution code and followed the evolutions of both the
mass donor and the accreting CO WD. The material that is lost as
optically thick wind is assumed to take away the specific orbital
angular momentum of the accreting WD, while the wind material that
is not accreted by the WD is assumed to take away the specific
orbital angular momentum of the donor star. For simplicity, we
chosen $M_{\rm WD}^{\rm i}=1.0$ $M_{\odot}$ and $1.1$ $M_{\odot}$,
$M_{\rm 2}^{\rm i}$ ranging from 0.8 $M_{\odot}$ to 5.6
$M_{\odot}$ step by 0.2 $M_{\odot}$ and initial orbital period
from $\log (P^{\rm i}/{\rm day})=1.5$ to 3.5 step by $0.1$, i.e.
the initial separation is from 55 $R_{\odot}$ to 1100 $R_{\odot}$.
Generally, the parameter space leading to SNe Ia deceases with the
initial WDs. Considering  the maximum CO WD mass is about 1.10
$M_{\odot}$ (Umeda et al. \cite{UME99}; Meng et al.
\cite{MENG08}), the possible parameter space leading to SNe Ia
here is an upper limit. In addition, the aim of choosing two
different initial WD masses is to show the potential dependence of
the final results on the initial WD mass.

In the calculations, we assume that the WD explodes as an SN Ia
when its mass reaches  a value close to the Chandrasekhar mass
limit, i.e., 1.378 $M_{\odot}$ (Nomoto, Thielemann \& Yokoi
\cite{NTY84}). According to the mass-transfer rate at that moment,
the supernovae may explode in the OTW stage, or after the wind
phase, where hydrogen-shell burning is stable (SSSs) or the mildly
unstable stage (RNe). At the moment of supernova explosion, the
companions  climb along the first-giant branch (FGB)\footnote{For
an FGB star, there is a degenerate helium core in the center of a
star, while there is a CO core in an AGB star.} or the asymptotic
giant branch (see Fig. \ref{m2mdot}). In Fig. \ref{m2mdot}, we
show the distribution of the final state of the symbiotic systems
leading to SNe Ia in a ($M_{\rm 2}^{\rm SN}$-$\log\dot{M}$) plane.
From this figure, we can see that the distribution of the final
state of the systems is clearly divided into two groups, i.e. the
top one is the AGB group and the bottom one is the FGB one.
Interestingly, only when the initial WD is massive enough, i.e.,
1.10 $M_{\odot}$, do the companions of the systems that produce
SNe Ia be AGB stars, as suggested by the X-ray observation of
Kepler's supernova remnant (Chiotellis et al. \cite{CHIOTELLIS12};
see also Chen et al. \cite{CHENXF11}). In addition, the mass-loss
rate from the system with the AGB companion is generally between
$10^{\rm -7}$ $M_{\odot}$ ${\rm yr}^{\rm -1}$ and $10^{\rm -6}$
$M_{\odot}$ ${\rm yr}^{\rm -1}$, while it is between lower than
$10^{\rm -8}$ $M_{\odot}$ ${\rm yr}^{\rm -1}$ and $10^{\rm -5}$
$M_{\odot}$ ${\rm yr}^{\rm -1}$ for  systems with  FGB companions.

At the moment of a supernova explosion, we record the status of the
binary system, e.g. the secondary mass, $M_{\rm 2}^{\rm SN}$, the
orbital separation, $a$, and the mass-loss rate from the system,
and so on. In the following sections, we show how we estimate the
luminosity of the X-Ray/radio (UV excess) from the interaction
between supernova ejecta and CSM (companion).

\subsection{The method to estimate the X-Ray flux from the interaction between supernova ejecta and CSM}\label{sect:3.4}
In our model, a large amount of material is lost from the system
by the optically thick wind or the tidally enhanced wind so that
CSM forms. Supernova ejecta interact with the CSM, and then X-ray
emission is expected. There are two models to estimate X-ray
luminosity. One is the thermal bremsstrahlung (TB) model, in which
X-ray emission is expected from the heated plasma (Immler et al.
\cite{IMMLER06}), the other is the inverse Compton (IC) model, in
which the X-ray emission from an SNe is dominated by the
scattering of photospheric optical photons from relativistic
electrons that were accelerated by the SN shock on a timescale of
weeks to a month after the explosion (Chevalier \& Fransson
\cite{CHEVALIER06}). Generally, the IC mechanism dominates the
X-ray flux from an SN Ia exploding in a low-CSM density
environment and the X-ray flux from this mechanism is very
sensitive to the observed bolometric light curve of an SN Ia (see
Margutti et al. \cite{MARGUTTI12} in their Appendix A). The X-ray
luminosity from the TB model in Immler et al. (\cite{IMMLER06}) is
generally consistent with that from the numerical simulation in
Dimitriadis et al. (\cite{DIMITRIADIS14}), i.e., from around
maximum light to 150 day later after an SNe Ia explosion, although
the model is very simple and some input parameters may be over- or
underestimated (see Dimitriadis et al. \cite{DIMITRIADIS14} in
details). This consistency is mainly because both Immler et al.
(\cite{IMMLER06}) and Dimitriadis et al. (\cite{DIMITRIADIS14})
only assumed emission from the reverse shock, as did also by
Hughes et al. (\cite{HUGHES07}). The estimated X-ray luminosity
from both models with a high given mass-loss rate may be
comparable with each other, but for a low mass-loss rate, the IC
mechanism dominates the X-ray emission. In addition, the X-ray
luminosity from the IC model decreases in much more quickly than
in the TB model (see the following text and Liu et al.
\cite{LIUZW15}). To account for both cases, we calculated the
X-ray luminosity for both models, respectively. It should be noted
here that we extrapolated the IC model in Margutti et al.
(\cite{MARGUTTI12}) to higher mass-loss rates, since Margutti et
al. (\cite{MARGUTTI12}) only dealt with single-scattering Compton
events, and  did not consider inverse Compton scattering on
thermal electrons (see also Lundqvist \& Fransson
\cite{LUNDQVIST88}), thereby limiting their analysis to low
mass-loss rates.

Regardless of the mechanism that produces X-ray emission, the X-ray
luminosity depends on a mass-loss rate. The measurement of the
X-ray emission may be helpful in constraining the mass-loss rate,
as well as the progenitor model since the CSM is generally expected
from the SD model. As indicated in Section \ref{sect:2}, no
conclusive X-ray detections are reported, which may only provide an
upper limit for mass-loss rate and seems not to uphold the
symbiotic systems as one progenitor channel of SNe Ia. Based on
the calculations from the TEW model here, we record the mass-loss
rate at the moment of supernova explosion and use the equation
from Immler et al. (\cite{IMMLER06}) and Margutti et al.
(\cite{MARGUTTI12}) to calculate the X-ray luminosity. For the TB
model, the X-ray luminosity is derived from
 \begin{equation}
L_{\rm X}=\frac{1}{\pi m^{\rm
2}}\Lambda(T)\left(\frac{\dot{M}}{v_{\rm w}}\right)^{\rm 2}(v_{\rm
s}t)^{\rm -1},\label{eq:lx}
  \end{equation}
where $L_{\rm X}$ is the X-ray luminosity, $m$ is the mean mass
per particle of $2.1\times10^{\rm -24}$ ${\rm g}$ for an H+He
plasma with solar composition, $t$ is the time after outburst,
$v_{\rm s}$ is the shock speed , which depends on the CSM density,
and the density slope of supernova ejecta (Chevalier
\cite{CHEVALIER82a}; Chomiuk et al. \cite{CHOMIUK12};
P\'{e}rez-Torres et al. \cite{PEREZ14}). But here for simplicity,
we set it as 10000 ${\rm km}$ ${\rm s}^{\rm -1}$, with the caveat
that this velocity can be much higher for a low density CSM. In
addition, $v_{\rm w}$ is the wind speed and $\Lambda(T)$ is the
cooling function: $\Lambda(T)=2.4\times10^{\rm -27}g_{\rm
ff}T_{\rm e}^{\rm 1/2}$, where $T_{\rm e}=1.36\times10^{\rm
9}(n-2)^{\rm -2}(\frac{v_{\rm s}}{10^{\rm 4} {\rm km} {\rm s}^{\rm
-1}})^{\rm 2}$ K is the electron temperature, and $g_{\rm ff}$ is
the free-free Gaunt factor. In this scenario, $n$ is the power
index of the outer density of supernova ejecta and is set to 10
(Chevalier \& Fransson \cite{CHEVALIER06}), and then, $T_{\rm
e}=2.13\times10^{\rm 7}$ K. Here, we assume that the reverse shock
dominates the X-ray flux as did in Immler et al.
(\cite{IMMLER06}), i.e., the X-ray luminosity from the reverse
shock is 30 times higher than that from the forward shock, and
then the total X-ray luminosity is the sum of the luminosity from
both forward and reverse shocks. Here, we assume the free-free
Gaunt factor to be 1 (Sutherland \cite{SUTHERLAND98}).

For the IC model, with a wind-like CSM structure, the X-ray
luminosity is derived from

 \begin{equation}
   \begin{array}{lc}
\frac{dL_{\rm x}}{d\nu}\approx2.1\times10^{\rm
-4}\left(\frac{\epsilon_{\rm e}}{0.1}\right)^{\rm
2}\left(\frac{M_{\rm ej}}{1.4M_{\odot}}\right)^{\rm
-0.93}\left(\frac{A}{{\rm g}\hspace{0.05cm}{\rm cm}^{\rm
-1}}\right)^{\rm
0.64}\\
\hspace{0.55cm}\times\left(\frac{E}{10^{\rm 51} {\rm
erg}}\right)^{\rm 1.29}\left(\frac{t}{\rm s}\right)^{\rm
-1.36}\left(\frac{L_{\rm bol}}{{\rm erg}\hspace{0.05cm}{\rm
s}^{\rm -1}}\right)\nu^{\rm -1},\label{eq:lxic}
  \end{array}
  \end{equation}
where the equation is obtained by assuming that electrons are
accelerated on the basis of a power-law distribution
$n(\gamma)\propto\gamma^{-p}$ with $p=3$ (Margutti et al.
\cite{MARGUTTI14}), and $\epsilon_{\rm e}=0.1$ is the fraction of
thermal energy in the shock that is used to accelerate the
electrons (Chevalier \& Fransson \cite{CHEVALIER06}), and
$A=\dot{M}/4\pi v_{\rm w}$. For simplicity, we set $E=10^{\rm 51}$
${\rm erg}$ and $M_{\rm ej}=1.4$ ${\rm M}_{\odot}$, which are
typical values for a normal SN Ia, and $L_{\rm bol}$ is set to be
$10^{\rm 43}$ ${\rm erg}$ ${\rm s}^{\rm -1}$, a typical value at
the maximum brightness epoch of a SN Ia.

Compared to  Equations (\ref{eq:lx}) and (\ref{eq:lxic}), we can
see that the TB mechanism is more sensitive to the mass-loss rate
and the wind velocity, while being less sensitive to an explosion
epoch than the IC model. For both models, $t$ is set at 20 days,
i.e., X-ray observations for these emissions are carried on around
maximum light. We assume a spherically symmetric mass loss, which
means that the lost material excavates a wind bubble around the
binary system. The CSM consists of two components. One is the mass
loss for the optically thick wind or helium flash from the surface
of the CO WD, where the wind velocity is $\sim1000$ ${\rm km}$
${\rm s}^{\rm -1}$ (Hachisu et al. \cite{HAC96}), the other is the
tidally enhanced wind from the secondary surface, where the wind
velocity is set as $\sim5$ ${\rm km}$ ${\rm s}^{\rm -1}$. We
calculate the X-ray luminosity from these two components according
to Equations (\ref{eq:lx}) and (\ref{eq:lxic}), respectively, and
the sum of the luminosity from the two components is taken as the
final X-ray luminosity.

However, please note that there are two connotative assumptions
in the above method: one is that the SNe Ia explode immediately
when the $M_{\rm WD}$ $=1.378$ $M_{\odot}$, i.e., we do not
consider the possible delay time of the explosion owing to the
rotation of the WD. The other is that the ejecta velocity is much
higher than the wind velocity. The second assumption usually holds,
while the first one may significantly affect the final results and
we will discuss it in Sections \ref{subs:5.2} and \ref{subs:5.3}.

\subsection{The method to estimate the radio flux from the interaction between supernova ejecta and CSM}\label{sect:3.5}
Radio radiation is also expected from the interaction between
supernova ejecta and the CSM. When a shockwave plows into the CSM,
it accelerates particles and amplifies the magnetic field, which
produces synchrotron emission that peaks in the cm band (Chevalier
\cite{CHEVALIER82a, CHEVALIER82b}). The early synchrotron signal
traces the CSM particle density, $n_{\rm CSM}$, on a radial scale
of $\leq1$ pc, in which the region is shaped by the final stage of
progenitor evolution, i.e., the final mass loss from the binary
system. So, the radio luminosity is mainly determined by the mass-loss
rate from the progenitor system. As a result, a measurement about the
radio radiation may obtain the information about the mass-loss
rate as well as the progenitor system. However, the VLA observations
on some archival SNe Ia may only provide an upper limit on the
radio emission as well as on the mass-loss rate (Panagia et al.
\cite{PANAGIA06}; Hancock et al. \cite{HANCOCK11}). According to
the TEW model here, we may obtain the mass-loss rate at the moment
of a supernova explosion. For simplicity, we adopt the parameterized
model in Panagia et al. (\cite{PANAGIA06}) and calculate the radio
luminosity based on the equation as followings:

 \begin{equation}
  \begin{array}{lc}
  \frac{L}{10^{\rm 26} {\rm erg} {\rm s}^{\rm -1}}=\Lambda\left(\frac{\dot{M}/10^{\rm -6}{\rm M}_{\odot} {\rm yr}^{\rm -1}}{v_{\rm w}/10 {\rm km}{\rm s}^{\rm -1}}\right)^{\rm 1.65}\\
  \hspace{1.25cm}\times\left(\frac{\nu}{\rm 5 GHz}\right)^{\rm -1.1}\left(\frac{t-t_{\rm 0}}{\rm day}\right)^{\rm -1.5}e^{-\tau_{\rm
  CSM}}.\label{eq:radio}
  \end{array}
  \end{equation}
Here, we set $\Lambda=1285$ and $\nu=5$ GHz. $t-t_{\rm 0}$ is set
to be a typical value of 20 days, i.e., around maximum light
(Panagia et al. \cite{PANAGIA06}). Equation (\ref{eq:radio}) is
for the forward shock, and generally, the radio emission from the
reverse shock is most likely significantly lower than that from
the forward shock. We then omit the radio emission from the
reverse shock and only take into account that from the forward
shock, which we proceed to calculate using Equation
(\ref{eq:radio}), where the forward shocked materials are from two
wind components, i.e., OTW and TEW. We still assume a spherically
symmetric mass loss, where the velocity of TEW, lost from the
surface of the WDs, is 1000 ${\rm km}$ ${\rm s}^{\rm -1}$ and the
TEW velocity is 5 ${\rm km}$ ${\rm s}^{\rm -1}$. We calculate the
radio luminosity from these two wind components according to
Equation (\ref{eq:radio}), respectively, and the sum of the
luminosity from the two wind components is taken as the final
radio luminosity. The external absorption depth, $\tau_{\rm CSM}$,
depends on the shock radius and the wind temperature (Weiler
\cite{WEILER86}; Lundqvist \& Fransson \cite{LUNDQVIST88}). For
simplicity, rather arbitrarily we assume  that $\tau_{\rm CSM}=0$,
which means that the results here are the conservative upper
limits (please see Equation \ref{eq:radio}). The connotative
assumptions in Section \ref{sect:3.4} may also exist, but we do
not repeat them again here.

\subsection{The method to estimate the optical/UV flux from the interaction of supernova ejecta and companion}\label{sect:3.6}
If SNe Ia are from the SD system, then after the explosion, a
large amount of material is ejected as a series of expanding
shells, which impact on the surface of the companion. The leading
edge of these expanding shells collides into the envelope of the
companion with a velocity of $V_{\rm SN}$ at $t_{\rm 0} = a/V_{\rm
SN}$, where $a$ is the orbital separation of the binary system at
the moment of the supernova explosion. These collisions may
produce an excess of optical/UV emission. The optical/UV emission
may exceed the radioactively powered luminosity of the supernova
for the first few days after the explosion and may reveal itself
by about 10\% of time (Kasen \cite{KASEN10}). However, the
detection of  this kind of signature is usually negative, which
constrains the companion in the SD model to be a MS star of less
than 6 $M_{\odot}$, and seems to disfavor an RG star undergoing
RLOF (Hayden et al. \cite{HAYDEN10}; Tucker \cite{TUCKER11};
Bianco et al. \cite{BIANCO11}; Brown et al. \cite{BROWN12b}). Even
taking the dependence of the luminosity on a viewing angle into
consideration, Brown et al. (\cite{BROWN12b}) may only rule out
the RG companion. Actually, the companion is usually less massive
than 2 $M_{\odot}$ (Han \cite{HAN08}; Hachisu et al. \cite{HKN08};
Meng \& Yang \cite{MENGYANG10c}) at the moment of a supernova
explosion. Therefore, it should be very difficult to detect the UV
excess from the WD + MS channel except that the companion is as
massive as 6 $M_{\odot}$ (Marion et al. \cite{MARION15}). The most
strict observational constraint is from SN 2011fe (Brown et al.
\cite{BROWN12a}), where the companion should be less massive than
1 $M_{\odot}$ assuming that the companion fills its Roche lobe.
Here, we use the equation (22)\footnote{Please keep in mind that
the equation (22) in Kasen (\cite{KASEN10}) was obtained by
assuming a RLOF. The secondaries in the symbiotic systems
considered here may not fill their Roche lobe and then the
collision luminosity based on the equation (22) in Kasen
(\cite{KASEN10}) could be overestimated in this paper. See
discussion in section \ref{subs:5.2}} in Kasen (\cite{KASEN10}) to
calculate the collision luminosity at three days after an
explosion as followings:

 \begin{equation}
L_{\rm c}=10^{\rm 43}a_{\rm 13}M_{\rm c}^{\rm 1/4}v_{\rm 9}^{\rm
7/4}\kappa_{\rm e}^{\rm -3/4}t_{\rm day}^{\rm -1/2} {\rm erg} {\rm
s}^{\rm -1},\label{lc}
  \end{equation}
where $t_{\rm day}$ is the time since the explosion measured in
days, and is set to be three days, which means the measurement of
the UV flux began 14-17 days before the maximum luminosity of SNe
Ia. $\kappa_{\rm e}=0.2$ ${\rm cm}^{\rm 2} {\rm g}^{\rm -1}$ and
$v_{\rm 9}=1$ is a transition velocity in units of $10^{\rm 9}
{\rm cm}$ ${\rm s}^{\rm -1}$ between inner and outer regions of
supernova ejecta. Here, $a_{\rm 13}$ is the orbital separation in
units of $10^{\rm 13}$ cm and $M_{\rm c}$ is the mass injected
into the companion in units of the Chandrasekhar mass. We set
$M_{\rm c}=\xi_{\rm v}\frac{\pi R_{\rm 2}^{\rm 2}}{4\pi a^{\rm
2}}=\xi_{\rm v}\frac{R_{\rm 2}^{\rm 2}}{4a^{\rm 2}}$, where
$R_{\rm 2}$ is the secondary radius and $\xi_{\rm v}$ is a
correction factor that depends on the ejecta velocity $V$ and
orbital separation $a$,

 \begin{equation}
\xi_{\rm v}=\frac{\int_{t_{\rm 0}}^{\rm t_{\rm day}}\rho_{\rm
SN}\cdot V\cdot dt\cdot 4\pi a^{\rm 2}}{M_{\rm Ch}},
  \end{equation}
where $V=a/t$. $V_{\rm SN}$ is set to be 10000 ${\rm km}$ ${\rm
s}^{\rm -1}$, which approximately corresponds to a total ejecta
kinetic energy of $1\times10^{\rm 51}$ ${\rm erg}$ ${\rm s}^{\rm
-1}$. Here, $\rho_{\rm SN}$ is the ejecta density at a distance
$a$ from the explosion center and is scaled as (Chugai
\cite{CHUGAI86}; Meng et al. \cite{MENGXC07})

 \begin{equation}
\rho_{\rm SN}=\frac{3M_{\rm Ch}}{4\pi a^{\rm 3}}\left(\frac{t_{\rm
0}}{t}\right)^{\rm 3}.
  \end{equation}
Then,
 \begin{equation}
\xi_{\rm v}=1-\left(\frac{t_{\rm 0}}{t_{\rm day}}\right)^{\rm 3}.
\label{xi}
  \end{equation}
Based on Equation (\ref{xi}), during the first or second day
after the supernova explosion, there may be no UV excess from the
interaction between supernova ejecta and companion if the orbital
separation is large enough. Here, we set $t_{\rm day}=3$ days,
which means that even for the largest orbital separation of a WD +
RG system leading to SN Ia, the supernova ejecta may collide into
the companion envelope at this time.

\section{Results}\label{sect:4}

   \begin{figure}
   \centering
   \includegraphics[width=60mm,height=80mm,angle=270.0]{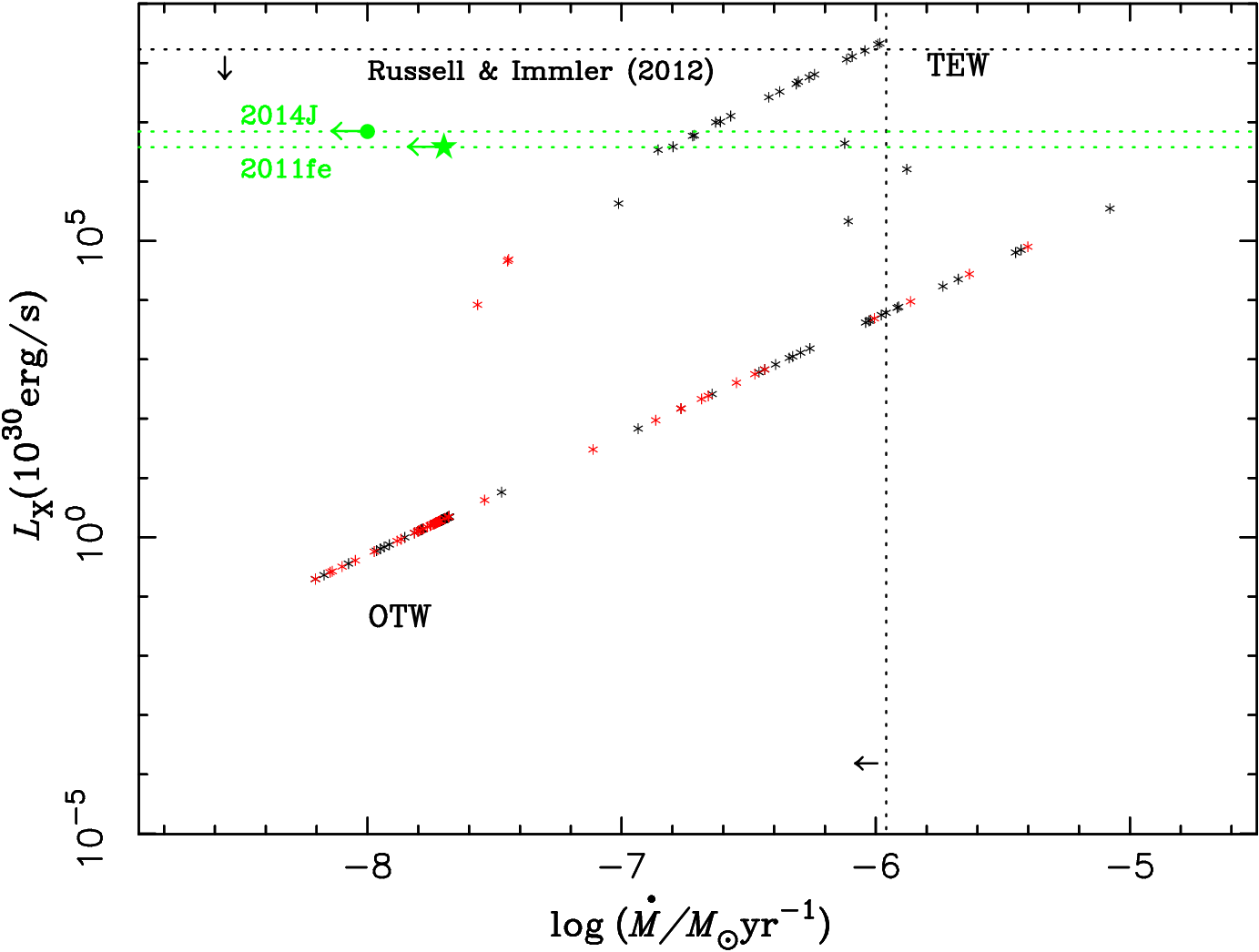}
   \includegraphics[width=60mm,height=80mm,angle=270.0]{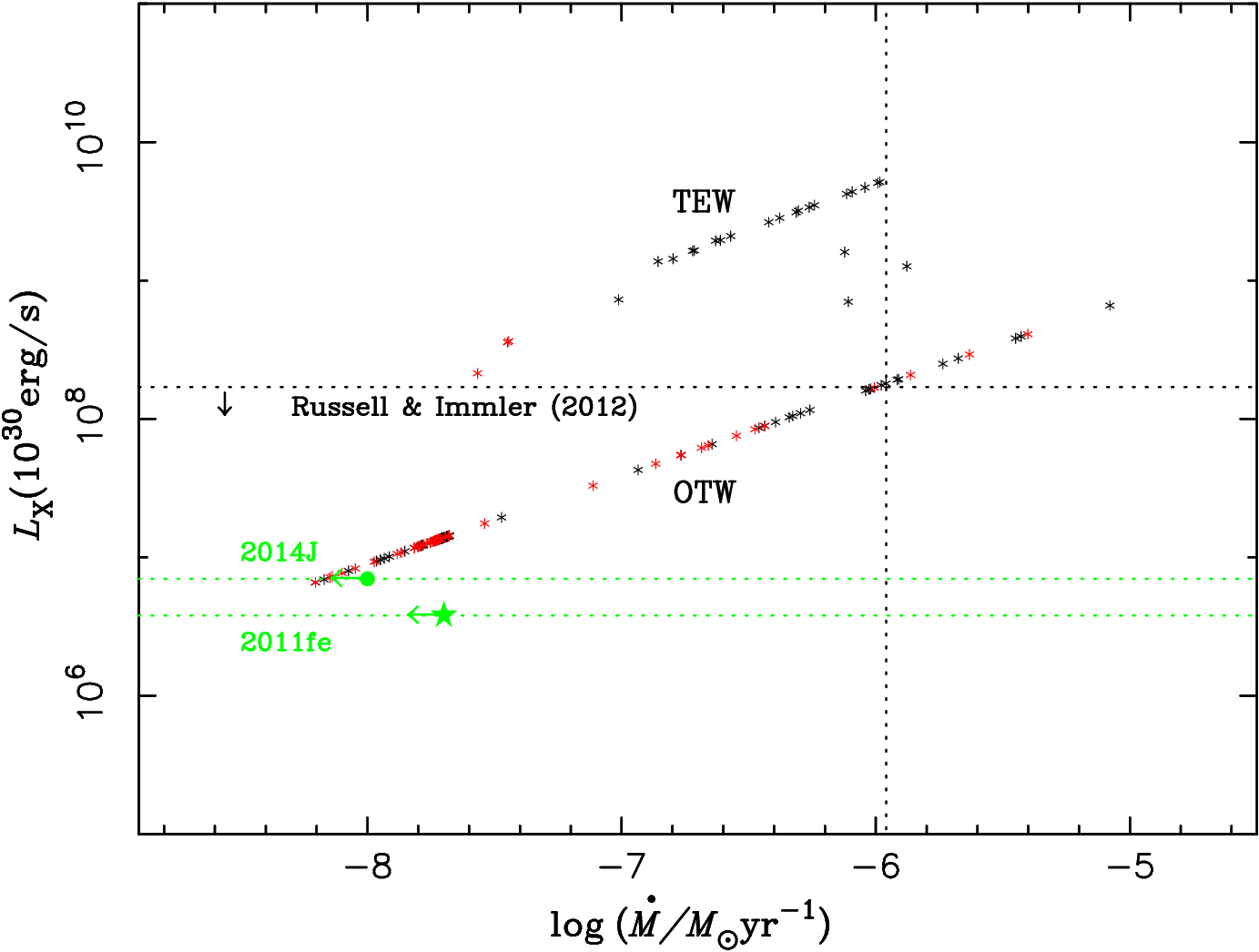}
   \caption{The X-ray luminosity on the basis of the different X-ray emission mechanisms
   from the interaction between supernova ejecta and CSM, as well as the total mass-loss rate from symbiotic systems
   at 20 days after supernova explosion.
 The black dotted lines are the observational constraints from Russell \& Immler (\cite{RUSSELL12}), while the green ones are
 for SNe 2011fe and 2014J in Margutti et al. (\cite{MARGUTTI12,MARGUTTI14}). The green filled star and circle show the upper limit of the
progenitor mass-loss rate for SNe 2011fe and 2014J, respectively,
assuming a wind velocity of 1000 km s$^{\rm -1}$ and the inverse
Compton mechanism dominating the X-ray flux based on the results
in Margutti et al. (\cite{MARGUTTI12,MARGUTTI14}). The X-ray
luminosity here is clearly divided into two sequences. The top
sequence indicates that the tidally enhanced wind (TEW) dominates
the mass loss, while the bottom one means that mass loss from the
surface of WDs as the optically thick wind (OTW) is dominant. The
red and black points are from the cases of $M_{\rm WD}^{\rm
i}=1.0M_{\odot}$ and $M_{\rm WD}^{\rm i}=1.1M_{\odot}$,
respectively. Upper: Based on the TB model with an assumption that
the luminosity from the reverse shock is 30 times higher than that
from the forward shock.
Bottom: Based on the IC model and extrapolating it to higher mass-loss rates.} \label{lxml}%
    \end{figure}
\subsection{X-ray emission from the interaction between supernova ejecta and CSM}\label{subs:4.1}
The X-ray luminosity estimated from the assumptions in Section
\ref{sect:3.4} is shown in Figure. \ref{lxml}, as well as the
total mass-loss rate from the binary systems. We also plot the
observational constraint from Russell \& Immler (\cite{RUSSELL12})
and two well observed SNe Ia, 2011fe and 2014J (Margutti et al.
\cite{MARGUTTI12,MARGUTTI14}). From the figure, we can see that
the maximum mass-loss rate at the moment of supernova explosion is
$\sim10^{\rm -5}$ $M_{\odot} {\rm yr}^{\rm -1}$, where the SNe Ia
explode in the OTW stage, and the minimum mass-loss rate is even
lower than $\sim10^{\rm -8}$ $M_{\odot} {\rm yr}^{\rm -1}$, where
the SNe Ia probably explode in the RN stage. In the figure, the
X-ray luminosity is clearly divided into two sequences, no matter
what the emission mechanism is. The top one indicates that the TEW
dominates the mass loss, while the bottom one indicates that mass
loss from the surface of WDs is dominant.

From Fig. \ref{lxml}, we can see that, for the TB model, the
maximum X-ray luminosity may be higher than $10^{\rm 38}$ ${\rm
erg}$ ${\rm s}^{\rm -1}$, even higher than the upper limit from
Russell \& Immler (\cite{RUSSELL12}), while the minimum value is
$\sim10^{\rm 29}$ ${\rm erg}$ ${\rm s}^{\rm -1}$, which is much
lower than those from any detection, i.e., for the TB model, almost
all the SNe Ia from the WD + RG systems fulfill the conservational
constraint from Russell \& Immler (\cite{RUSSELL12}). However, for
the IC model, the maximum X-ray luminosity is close to $10^{\rm
40}$ ${\rm erg}$ ${\rm s}^{\rm -1}$, and the luminosity for almost
all the systems is higher than $10^{\rm 37}$ ${\rm erg}$ ${\rm
s}^{\rm -1}$. Especially, the X-ray luminosity from all the WD +
AGB models is higher than the upper limit from Russell \& Immler
(\cite{RUSSELL12}), but a large amount of WD + FGB models still
fulfill this constraint. Compared with TB and IC models, for a
lower mass-loss rate, the X-ray luminosity from the IC model is
higher than that from the TB model by a factor of $\sim10^{\rm
7}$, but for a higher mass-loss rate, the X-ray luminosity from
the IC model is higher than that from the TB model by only a
factor of dozens, i.e., the difference between the two models
decreases. However, for a high mass-loss rate, the X-ray
luminosity from these two models would very likely be comparable,
especially regarding the material from the TEW. Therefore, the
significant difference for high mass-loss rates between the two
models shown in Fig. \ref{lxml} implies that the IC model proposed
in Margutti et al. (\cite{MARGUTTI12}) is not suitable for direct
extrapolation to high-mass loss rate, since multiple scattering
and inverse Compton scattering on thermal electrons were not
accounted for in the model (Lundqvist \& Fransson
\cite{LUNDQVIST88}).

For the strict constraints from two well observed SNe Ia, 2011fe
and 2014J, the WD + AGB systems are very likely excluded, whatever
the emission mechanism is. However, for the TB model, the X-ray
luminosity from many WD + FGB models fulfill the observational
limit of \emph{Chandra} ($3.8\times10^{\rm 36}$ ${\rm erg}$ ${\rm
s}^{\rm -1}$, Margutti et al. \cite{MARGUTTI12}), while for the IC
model, the X-ray observations of 2011fe and 2014J almost ruled out
all the WD + FGB models. As a result, we confirm that SN 2011fe
and 2014J are unlikely to come from the WD + AGB system on the
basis of their low progenitor mass-loss rate.

So, at present, it seems that the X-ray observation has the
ability to rule out the WD +AGB systems as the progenitor of SNe
Ia if the tidally-enhanced wind velocity is not much higher than 5
km s$^{\rm -1}$, as adopted here, and the IC mechanism dominates
the X-ray emission from low-density CSM. On the other hand, the WD
+ FGB systems cannot be excluded completely, and a definite
conclusion very much depends on the X-ray emission mechanism and
on the distance of the SNe Ia. The wind velocity from MS stars is
much higher than that from RG ones, which makes it difficult to
detect X-ray from an SN Ia arising from the WD + MS channel.
Generally, there is no CSM around a DD system before a supernova
explosion, and therefore, the X-ray observation may give a
meaningful constraint to the WD + RG channel if the SNe Ia are as
near as SNe 2011fe and 2014J. In any case,  X-ray flux is more
likely to be detected from the WD + AGB system, rather than the WD
+ FGB system, based on the results shown in Figs. \ref{m2mdot} and
\ref{lxml}, since the material is more likely to be lost as the
tidally enhanced wind with a low velocity for the WD + AGB system.
We note that only when the initial WD mass is high enough, i.e.
$M_{\rm WD}^{\rm i} = 1.1 M_{\odot}$, may the WD + AGB systems be
the progenitors of SNe Ia.

   \begin{figure}
   \centering
   \includegraphics[width=60mm,height=80mm,angle=270.0]{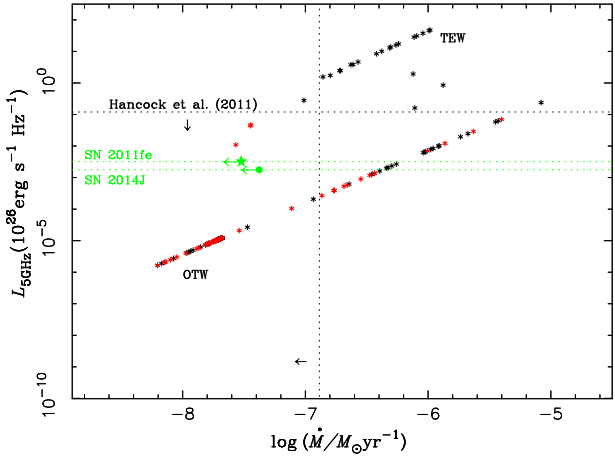}
   \caption{The radio luminosity of 5 GHz from the interaction between supernova ejecta and CSM, as well as the total mass-loss rate from the symbiotic system, 20 days after a supernova explosion.
 The black dotted lines are the observational constraints from Hancock et al. (\cite{HANCOCK11}), while the green dotted lines are for SNe 2011fe and 2014J.
 The data for 2011fe is the stacked EVLA radio luminosity on an averaged date of 9.1 days
after explosion and an average baseband of 5.9 GHz (Chomiuk et al.
\cite{CHOMIUK12}). The data for 2014J is from the observation of
5.5 GHz, 8.2 days after the explosion, which is the most stringent
constraint among the observations in P\'{e}rez-Torres et al.
(\cite{PEREZ14}). The green-filled star and circle show the upper
limit of the progenitor mass-loss rate for SNe 2011fe and 2014J,
respectively, assuming a wind velocity of 1000 km s$^{\rm -1}$.
The radio luminosity here is clearly divided into two sequences.
The top sequence indicates that the tidally enhanced wind
dominates the mass loss, while the bottom one means that mass loss
from the surface of WDs is dominant. The red and black points are
from the cases of $M_{\rm WD}^{\rm i}=1.0M_{\odot}$ and $M_{\rm
WD}^{\rm i}=1.1M_{\odot}$,
   respectively.} \label{radio}%
    \end{figure}

\subsection{Radio emission from the interaction between supernova ejecta and CSM}\label{subs:4.2}
Figure \ref{radio} shows the radio luminosity that is estimated
from the assumptions in Section \ref{sect:3.5}, as well as the
total mass-loss rate from the binary systems. The observational
constraints from Hancock et al. (\cite{HANCOCK11}) and two
well-observed SNe Ia, 2011fe, and 2014J (Chomiuk et al.
\cite{CHOMIUK12}; P\'{e}rez-Torres et al. \cite{PEREZ14}) are also
plotted in the figure. The results here are similar to those in
Section \ref{subs:4.1}. The radio luminosity is also divided into
two sequences. The top one is derived from the tidally enhanced
mass loss from the secondary, and the bottom one is derived from
the mass loss from the surface of the CO WD by the optically thick
wind or helium flash. In addition, the maximum radio luminosity is
less than $10^{\rm 28}$ ${\rm erg}$ ${\rm s}^{\rm -1}$ ${\rm
Hz}^{\rm -1}$, which is higher than the upper limit obtained from
VLA observations ($1.2\times10^{\rm 25} {\rm erg}$ ${\rm s}^{\rm
-1}$ ${\rm Hz}^{\rm -1}$, Hancock et al. \cite{HANCOCK11}). But
almost all of them, whose radio luminosity is higher than the
observational limit, are from the WD + AGB systems. Such WD + AGB
systems are relatively rare according to the initial mass
function, which is consistent with the non-detection results in
Panagia et al. (\cite{PANAGIA06}) and Hancock et al.
(\cite{HANCOCK11}). However, there is still a large parameter
space fulfilling the constraint from the radio observations, which
are from the WD + FGB systems. The WD + FGB systems are the
dominant channel to produce SNe Ia via the symbiotic channel (see
also Fig. 2 in Chen et al. \cite{CHENXF11}). Even for SN 2011fe,
which is believed to be exploding in a very clean environment
(Patat et al. \cite{PATAT13}), many models still fulfill the EVLA
constraints ($3.2\times10^{\rm 23}$ ${\rm erg}$ ${\rm s}^{\rm -1}$
${\rm Hz}^{\rm -1}$). Although the origin of the CSM of SN 2014J
is controversial (Graham et al. \cite{GRAHAM15}; Soker
\cite{SOKER15}; Maeda et al. \cite{MAEDA16}), many models here
also fulfill its radio constraint. So, it seems impossible for
radio research to rule our WD + FGB systems as the progenitor of
SNe Ia, even for SNe Ia as near as SN 2011fe and 2014J, but the
radio flux may be detected by present observation apparatus if the
SNe Ia are derived from the WD + AGB systems and the
tidly-enhanced wind velocity is not much higher than that 5 km
s$^{\rm -1}$ that is adopted here. As a result, we confirm that it
it is unlikely that SNe 2011fe and 2014J are from the WD + AGB
systems owing to their low radio luminosity.

   \begin{figure}
   \centering
   \includegraphics[width=60mm,height=80mm,angle=270.0]{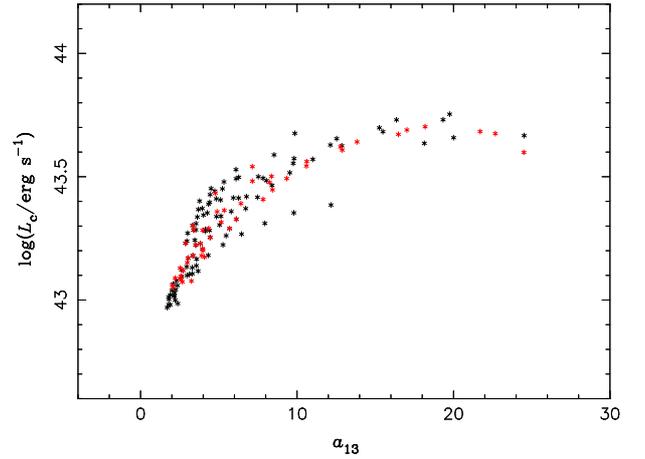}
   \caption{Collision luminosity at $t_{\rm day}=3$ days after supernova explosion from the interaction between supernova ejecta and companion, as
   well as
 the orbital separation in unit of $10^{\rm 13}$ cm. The red and black points are from the cases of $M_{\rm WD}^{\rm i}=1.0M_{\odot}$ and $M_{\rm WD}^{\rm i}=1.1M_{\odot}$,
   respectively.}
   \label{collsion}%
    \end{figure}
\subsection{The optical/UV flux from the interaction of supernova ejecta and companion}\label{subs:4.3}
Following on from the descriptions in Section \ref{sect:3.6}, we
obtain the collision luminosity shown in Figure \ref{collsion}.
For most cases in the symbiotic systems considered here, the
excess UV luminosity is between $10^{\rm 43}$ ${\rm erg}$ ${\rm
s}^{\rm -1}$ and $10^{\rm 44}$ ${\rm erg}$ ${\rm s}^{\rm -1}$,
similar to luminosity of Kasen (\cite{KASEN10}). We note, however,
that we have not considered the effect of the view angle, i.e. the
luminosity that results from the collision is obtained for the
viewing angles looking down on the collision region
($\theta=0^{\rm \circ}$). However, if the companion of a SN Ia is
a RG star which fills its Roche lobe, a very early observation for
the SN Ia could detect the UV excess flux if  the SN Ia is viewed
along the collision region. In addition, the UV excess flux is not
sensitive to the initial WD mass and to whether the companion is
an AGB star or a FGB. So, compared with the X-ray and radio
observations, the probability of detecting the UV excess flux is
relatively high, which is consistent with the results that there
are a few positive reports for the UV excess flux-detection while
there are negative reports for the X-ray/radio detection.

\section{Discussions}\label{sect:5}
\subsection{X-ray/radio emission}\label{subs:5.1}
As discussed in Sections \ref{sect:2}, \ref{subs:4.1}, and
\ref{subs:4.2}, the X-ray and radio observations may provide
meaningful constraints on the mass-loss rate from progenitor
systems, but only have an ability to rule out WD + AGB systems as
the progenitor of SNe Ia for the present observation apparatus.
For the SNe Ia from the WD + FGB systems, X-ray flux is generally
expected if the SNe Ia are as near as SN 2011fe and 2014J, while
radio emission may be difficult to detect, even if the SNe Ia are
as near as SN 2011fe and 2014J.

In fact, the real physics for X-ray and radio emission may be more
complex than the simple methods adopted here, such as the radio
emission, as well as the X-ray emission, is sensitive to varying
microphysics parameters and at which epoch a supernova is observed
(P\'{e}rez-Torres et al. \cite{PEREZ14}; Margutti et al.
\cite{MARGUTTI12}). For the simple analytic methods adopted here,
two key factors mainly determine the fluxes of X-ray and radio
emission. One is the mass-loss rate. For the symbiotic channel
considered here, a large amount of materials are lost from the
secondary before the supernova explosion, and then the mass-loss
rate at the moment of a supernova explosion may not be as high as
that expected from a single RG star. The other is the wind
velocity. For WD + MS systems, the optically thick wind dominates
the mass loss (Han \& Podsiadlowski \cite{HAN04}; Hachisu et al.
\cite{HKN08}; Meng, Chen \& Han \cite{MENG09}; Meng \& Yang
\cite{MENGYANG10}). The wind velocity for the optically thick wind
may be as high as $\sim1000$ ${\rm km}$ ${\rm s}^{\rm -1}$, which
makes it very difficult to detect the interaction between the
supernova ejecta and the CSM from the optically thick wind by the
radio observation. In addition, for the SD model, the WD + MS
channel is the main contributor to SNe Ia, and even if the
material is lost from the surface of the MS companions, instead of
the surface of the WDs, the lost materials are still difficult to
 detect by radio observation owing to their relatively high wind
velocity. Fortunately, X-ray observation could provide a powerful
tool to detect such low-density CSM if the SNe Ia are near enough.
For the material lost from the RG star, we note that
the tidally-enhanced wind velocity adopted here is a lower limit,
which could be much lower than a real value. So, the maximum
luminosity of X-ray and radio emission here would represent an
upper limit.

Statistically, the progenitor systems leading to SNe Ia are more
likely to be RNe at the moment of supernova explosion (Meng \&
Yang \cite{MENGYANG11}), in which the CSM structure may be shaped
by a series of nova explosions. The X-ray luminosity from the
interaction between the supernova ejecta and the shaped CSM is at
least one order of magnitude less than that of an SN evolving in a
wind bubble (Dimitriadis et al. \cite{DIMITRIADIS14}). Therefore,
the X-ray luminosity from most SNe Ia could be much smaller than
that shown in Russell \& Immler (\cite{RUSSELL12}), if only the
thermal X-ray emission is considered. For low CSM densities,
however, this reduction could be easily offset by inverse Compton
scattering.

In particular, radio emission from SN 2002ic was not detected by
the VLA (Berger et al. \cite{BERGER03}; Stockdale et al.
\cite{STOCKDALE03}). SN 2002ic is the first SN Ia to definitely
show a CSM signal (Hamuy et al. \cite{HAM03}), where the amount of
CSM is very massive, i.e. $0.5$ $M_{\odot}$ $\sim$ 6 $M_{\odot}$
(Wang et al. \cite{WANGL04}; Chugai \& Yungelson \cite{CHUGAI04};
Uenishi et al.  \cite{UENISHI04}; Kotak et al. \cite{KOTAK04}).
Moreover, the radio emission from another twin of SN 2002ic (SN
2005gj, Aldering et al. \cite{ALDERING06}) was also not detected
by the VLA (Soderberg \& Frail \cite{SODERBERG05}). Although the
non-detection of radio flux was not completely unexpected from SN
2002ic (Chugai et al. \cite{CHUGAI04b}), non-detection results
from both SNe 2002ic and 2005gj could indicate that the mechanism
that was successfully used for SNe Ib/c might not work for SNe Ia.

In Sections \ref{subs:4.1} and \ref{subs:4.2}, we assume that the
mass loss is spherically symmetric. The real structure may be more
complex than this simple structure. For example, the tidally
enhanced wind concentrates near the orbital plane, while the wind
from the surface of the WDs is mainly along the direction
that is perpendicular to the orbital plane. As an example,
three-dimensional modelling of RS Oph has shown that the CSM is
expected to be concentrated in the binary orbital plane, which
suggests that the probability of detecting CSM is also strongly
dependent on the viewing angle (Mohamed et al. \cite{MOHAMED13}).
However, the complex structure may only change the X-ray/radio
luminosity estimated here by a factor of 2, and the effect of the
asymmetry may be partly counteracted by a relatively high shock
velocity that was adopted in this paper. Therefore, the discussions here
still hold, at least to first order.

In addition, based on the results in this paper and the present
detection limit, the expected X-ray/radio flux detection is more
likely from the WD + AGB systems, in which the initial WD mass
must be massive, i.e. $M_{\rm WD}^{\rm in}=1.1 M_{\odot}$ . But,
compared with the WD + FGB systems, these systems are relatively
rare owing to the initial mass function, which increases the
difficulty to detect the X-ray/radio flux predicted from the
symbiotic systems. Even for WD + FGB systems, their contribution
to all SNe Ia may only be several per cent (Han \& podsiadlowski
\cite{HAN04}; Ruiter et al. \cite{RUITER09}). Furthermore, we have
not considered the possible delay time between the moment of
$M_{\rm WD}=1.378$ $M_{\odot}$ and supernova explosion for the
rapid rotation of the accreting WDs (Justham \cite{JUSTHAM11}; Di
Stefano et al. \cite{DISTEFANO11}), which may significantly clean
the environment of a part of the SNe Ia.

\subsection{Optical/UV Flux}\label{subs:5.2}
As shown in Section \ref{subs:4.3}, the optical/UV flux from the
impact between supernova ejecta and its RG companion is very high,
even higher than the radioactively powered luminosity of the
supernova. Therefore, the optical/UV emission should be detected.
However, the report of the detection of the UV excess was usually
negative, except for SN 2009ig, 2011de, 2012cg, and iPTF14atg
(Foley et al. \cite{FOLEY12b}; Brown \cite{BROWN14}; Marion et al.
\cite{MARION15}; Cao et al. \cite{CAOY15}). This could be a
natural result in statistics. Kasen (\cite{KASEN10}) predicted
that about 10\% of all SN Ia have a favourable viewing angle with
strong detectable emission in the ultraviolet and B optical bands,
but this does not mean that we can detect the UV excess in the
light curve of about 10\% of all SN Ia. In observations, the UV
luminosity cannot be detected for a MS star with a mass smaller
than 6 $M_{\odot}$ (in which case SN 2012cg may be a wonderful candidate, Marion et al. \cite{MARION15}). Actually, the companion
is usually less massive than 2 $M_{\odot}$ at the moment of
a supernova explosion (Han \cite{HAN08}; Hachisu et al.
\cite{HKN08}; Meng \& Yang \cite{MENGYANG10c}). The UV excess from
the WD + MS channel is therefore almost undetected. For the
SNe Ia from the WD + RG channel in which the RG is experiencing
RLOF, even if the birth rate is very uncertain, their
contribution to all SNe Ia must be lower than 1\% $\sim$ 10\%
(Branch et al. \cite{BRANCH95}; Meng \& Yang \cite{MENGYANG10};
Wang, Li \& Han \cite{WANGB10}). Although the parameter space for
SNe Ia is enlarged for the tidally enhanced stellar wind (Chen et
al. \cite{CHENXF11}), the contribution of the SNe Ia from the wind
accretion symbiotic systems is still in a magnitude of a few
percent (see discussion in Han \& Podsiadlowski \cite{HAN04}). So,
the 10\% predicted by Kasen (\cite{KASEN10}) should be multiplied
by a factor of 0.01 $\sim$ 0.1 for SNe Ia from WD + RG channel,
in which case, in about 100 $\sim$ 1000 SNe Ia, we may have an
opportunity to see the UV excess from 1 SNe Ia. Bearing in
mind that, for most SNe Ia from the symbiotic systems, the
luminosity obtained here could be overestimated since Equation
(\ref{lc}) is deduced from a RLOF assumption. As such, the number 1
in 100 $\sim$ 1000 SNe Ia is a conservative upper limit, which is
consistent with a number of detections within several hundred SNe
Ia.

In addition, Justham (\cite{JUSTHAM11}) and Di Stefano et al.
(\cite{DISTEFANO11}) have suggested that the radius of the donor
star in the SD model may  shrink remarkably prior to a supernova
explosion for the exhaustion of its hydrogen-rich envelope owing
to a long spin-down time of the rapidly rotating WD. Therefore,
the cross-section of supernova ejecta would become smaller and the
optical/UV luminosity could be much smaller than that from the
Roche lobe model considered in Kasen (\cite{KASEN10}) (see also in
Hachisu et al. \cite{HKSN12}). For example, Justham
(\cite{JUSTHAM11}) have shown that the companion radius may shrink
from about 100 $R_{\rm \odot}$ to a value lower than 1 $R_{\rm
\odot}$. Assuming that the angle of the oblique shock front
relative to the flow direction also decreases by factor of 100,
then the UV flux presented in Fig. \ref{collsion} is overestimated
by at least five orders of magnitude, which is much lower than
that from the radioactively powered luminosity of the supernovae.
However, it must be mentioned that the spin-down timescale of the
WDs is quite uncertain (see discussions in Meng \& Podsiadlowski
\cite{MENGPOD13}) and, based on the discussion in Cao et al.
(\cite{CAOY15}), the radius of the companion of iPTF14atg seems
not to  shrink significantly, as suggested in Justham
(\cite{JUSTHAM11}). Furthermore, Ganeshalingam
(\cite{GANESHALINGAM11}) noticed that a substantial degree of
degeneracy exists between the adopted power-law index of the SN
light-curve template, the rise time, and the amount of shock
emission required to match the data, which makes it very difficult
to distinguish the UV excess from the UV flux of SNe Ia. The
discovery of Ganeshalingam (\cite{GANESHALINGAM11}) could indicate
that it is very difficult to detect this type of UV flux, except
that the UV excess is very significant, such as that from WD + RG
systems. Moreover, Kutsuna \& Shigeyama (\cite{KUTSUNA15})
recently found that the UV flux could not be detected from the
binary systems with a separation less than $2\times10^{\rm 13}$
cm. As shown in Fig. \ref{collsion}, a lot of systems that lead to
SNe Ia have a separation that is less than this separation limit.
However, for a very large separation, the supernova ejecta may not
collide into the companion within one or two days after the
supernova explosion (see Equation \ref{xi}), which is the best
time to detect the UV excess (e.g. Fig. 2 in Kasen
\cite{KASEN10}). This situation reduces the opportunity to detect
the UV excess further. For the discussion above, the rare report
of the UV excess from the interaction between supernova ejecta and
companion is a natural result in both physics and statistics, and
we can expect one in several hundred SNe Ia to show this kind of
UV excess in their early light curves.

   \begin{figure}
   \centering
   \includegraphics[width=60mm,height=80mm,angle=270.0]{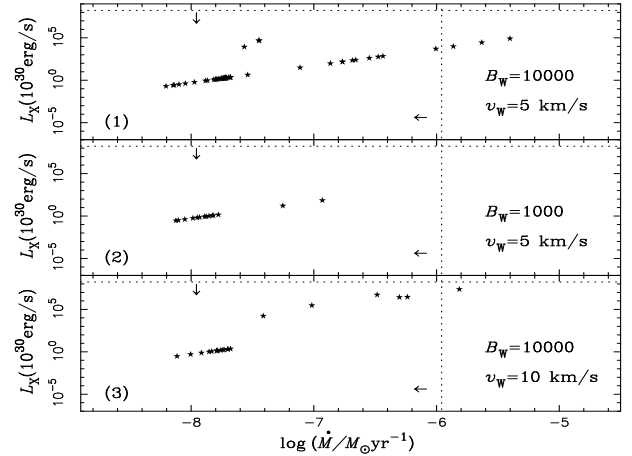}
   \caption{Similar to the upper panel in Fig. \ref{lxml} but for various $B_{\rm W}$ and $v_{\rm W}$ as shown in this figure, where $M_{\rm WD}^{\rm i}=1.00M_{\odot}$.
 The dotted lines are the observational constraint from Russell \& Immler (\cite{RUSSELL12}).}
   \label{mixlx}%
    \end{figure}
\subsection{Uncertainties}\label{subs:5.3}
In this paper, we assume that when $M_{\rm WD}=1.378 M_{\odot}$,
the WD explodes as an SNe Ia, and then we recorded the parameters
to estimate the X-ray/radio/optical/UV flux at the moment of
$M_{\rm WD}=1.378 M_{\odot}$. If the rapid rotation of the WD is
considered, the WD may increase its mass to 2.0 $M_{\odot}$ (Yoon
\& Langer \cite{YOON04,YOON05}; Justham \cite{JUSTHAM11}; Hachisu
et al. \cite{HKSN12}) and the parameters, e.g. mass-loss rate and
the companion radius, could be smaller than the value used here.
Meng \& Podsiadlowski (\cite{MENGPOD13}) estimate that the delay
time from the moment of $M_{\rm WD}=1.378 M_{\odot}$ to the
supernova explosion is less than a few $10^{\rm 7}$ yr. If the
delay time is as long as $10^{\rm 7}$ yr, all FGB companions may
have enough time to become He WDs or horizontal branch stars, and
most of the AGB companions to become CO WDs. Therefore, the estimated
X-ray/radio/optical/UV flux here should be taken as an upper
limit.

Actually, the WD + RG channel discussed in this paper is strongly
dependent on two poorly known parameters, i.e. $B_{\rm W}$ and
$v_{\rm W}$, which may significantly change the parameter space
leading to SNe Ia (Chen et al. \cite{CHENXF11}). Here, we choose
two values, e.g. $B_{\rm W}=1000$ and $v_{\rm W}=10$ ${\rm km}
{\rm s}^{\rm -1}$ to test their effect on the results in this
paper. We only show the effect on X-ray emission based on
the TB model and the discussions here are also applicable to
radio/UV emissions and X-ray emission from the IC mechanism, since
these discussions are not specific to the emission mechanism
involved.

In Figure \ref{mixlx}, we show the X-ray luminosity from the TB
mechanism and the total mass-loss rate from the symbiotic system
for different $B_{\rm W}$ and $v_{\rm W}$, where $M_{\rm WD}^{\rm
i}=1.00M_{\odot}$. A clear trend in the figure is that the number
of models leading to SNe Ia decreases with decreasing $B_{\rm W}$
and increasing $v_{\rm W}$, which is consistent with the discovery
in Chen et al. (\cite{CHENXF11}). It is clearly shown in the
figure that the basic results here still hold no matter what
$B_{\rm W}$ and $v_{\rm W}$ are. In addition, the maximum X-ray
luminosity becomes lower for a low initial WD mass (see Figure
\ref{lxml} and Panel 1 in Figure \ref{mixlx}). Generally, the
$M_{\rm WD}^{\rm i}$ is more likely to be around 0.8 $M_{\odot}$
owing to the effect of the initial mass function, although the
parameter space that leads to SNe Ia for $M_{\rm WD}^{\rm i}$ = $0.8$
$M_{\odot}$ is much smaller than that for $M_{\rm WD}^{\rm i}$ =
$1.1$ $M_{\odot}$ (Meng, Chen \& Han \cite{MENG09}; Meng \& Yang
\cite{MENGYANG10}; Chen et al. \cite{CHENXF11}). As a result, for
most SNe Ia, the real X-ray flux would be lower than the maximum
value estimated here.

For this study, we did not carry out a binary population synthesis
(BPS) study but just show a possible range for the luminosity of
the X-ray/radio/UV emission from the interaction between supernova
ejecta and the CSM/secondary. Since the initial mass of the CO WDs
is generally less massive than 1.1 $M_{\odot}$ (Umeda et al.
\cite{UME99}; Meng et al. \cite{MENG08}), e.g, generally the
initial WD mass is mainly between 0.8 $M_{\odot}$ and 1.0
$M_{\odot}$ (Chen et al. \cite{CHENXF11}), and considering that
the maximum luminosity of the X-ray/radio/optical/UV emission
decreases with a reduction in  the initial mass as presented in
Figure \ref{lxml} and Panel 1 in Figure \ref{mixlx}, the
luminosity range shown here is a conservative upper limit and thus
the basic results obtained in this paper are reasonable.

However, we emphasise that the emission mechanisms for X-ray/radio
are very complex and that the analytic methods adopted here are
very simple. In fact, for IC and radio emission, there are several
significant uncertainties, e.g. the $\epsilon$ parameters in
Margutti et al. (\cite{MARGUTTI12}) and P\'{e}rez-Torres et al.
(\cite{PEREZ14}), which may change the X-ray/radio luminosity by
as much as a factor of $\sim10$, none of which have been accounted
for in our study. For example, $\epsilon_{\rm e}=0.3$ in Equation
(\ref{eq:lxic}) is possible (Chevalier \& Fransson
\cite{CHEVALIER06}), in which case the X-ray luminosity from the
TB mechanism would become comparable to that from the IC mechanism
at a high $\dot{M}/v_{\rm w}$ value. In any case, the IC mechanism
dominates the X-ray emission for low wind densities, and the X-ray
luminosity will be significantly underestimated at low
$\dot{M}/v_{\rm w}$ values when only the TB mechanism is
considered. However, the IC model adopted here does not take the
effects of multiple scattering and inverse Compton scattering on
thermal electrons into consideration, which will become important
at high $\dot{M}/v_{\rm w}$. Although the ratio of thermal to
non-thermal inverse Compton scattering is complicated to estimate,
the situation in which the inverse Compton scattering on the
thermal electrons becomes significant may be similar to that for
the core-collapse SN 1993J, with only a somewhat larger
contribution from the inverse Compton scattering, compared to that
of thermal emission (Fransson et al. \cite{FRANSSON96}).
Therefore, regarding the X-ray luminosities obtained in this
paper, those from the TB mechanism at the high $\dot{M}/v_{\rm w}$
end and those from the IC mechanism at the low $\dot{M}/v_{\rm w}$
end can better approximate actual observations, such as those
discussed in Chevalier et al. (\cite{CHEVALIER06b}) (see also
Immler et al. \cite{IMMLER07}).

\section{Conclusions}\label{sect:6}
In summary, incorporating the effect of tidally enhanced wind on
the evolution of WD binaries into Eggleton's stellar evolution
code, and also including  the prescription of Hachisu et al.
(\cite{HAC99a}) for mass-accretion of CO WD, we performed a series
of binary evolution calculations for WD + RG systems. Assuming
that an SN Ia occurs when $M_{\rm WD}=1.378$ $M_{\odot}$, we found
that the mass-loss rate at the moment of a supernova explosion is
generally from a value lower than $10^{\rm -8}$ $M_{\rm \odot}$
${\rm yr}^{\rm -1}$ to $10^{\rm -5}$ $M_{\rm \odot}$ ${\rm
yr}^{\rm -1}$. For an SN Ia from a WD + AGB system, the mass loss
is dominated by TEW, while dominated by OTW for an SN from a WD +
FGB system. Based on the calculations here, we estimated the
X-ray/radio (UV excess) luminosity from the interaction between
supernova ejecta and CSM (secondary) using some standard published
models. We found that if an SN Ia is derived from a WD + AGB
system and the wind velocity is not much higher than the 5 ${\rm
km}$ ${\rm s}^{\rm -1}$ adopted in this paper, it may be detected
by both X-ray and radio observations. For an SN Ia from a WD + FGB
system, it is generally difficult to detect radio emission from
the supernova since the materials lost from the system as the OTW
usually move at high velocity, even for SNe Ia as near as SN
2011fe and 2014J, but X-ray observation may provide a meaningful
constraint, i.e., the X-ray emission may be detected from an SNe
Ia that is derived from a symbiotic system if it is near enough,
in which the inverse Compton scattering dominates the X-ray
emission. For the two well observed SNe Ia, 2011fe and 2014J,
almost all the symbiotic systems are excluded by X-ray
observations, while WD + FGB systems may not be excluded
completely through radio observation. As a result, we confirm that
SNe 2011fe and 2014J very unlikely arise from WD + AGB systems.
The UV excess luminosity from the interaction between supernova
ejecta and RG secondaries is usually comparable to the
radioactively powered luminosity of the supernova for the first
few days after a supernova explosion if the companion fills its
Roche lobe, no matter whether the companion is an AGB star or an
FGB star. However, we can only expect one among hundreds of SNe Ia
to show the excess UV emission. So, the rare reporting of UV
excess that is derived from the interaction between SN ejecta and
companion is a natural result in both physics and statistics.

\begin{acknowledgements}
We are very grateful to the anonymous referee for his/her
constructive suggestions that greatly improved this manuscript.
This work was partly supported by NSFC (11473063,11522327,
11390374, 11521303), CAS ``Light of West China'' Program, CAS (No.
KJZD-EW-M06-01) and Key Laboratory for the Structure and Evolution
of Celestial Objects, Chinese Academy of Sciences. Z.H. thanks the
support by the Strategic Priority Research Program ``The Emergence
of Cosmological Structures'' of the Chinese Academy of Sciences,
Grant No. XDB09010202, and Science and Technology Innovation
Talent Programme of the Yunnan Province (Grant No. 2013HA005).
\end{acknowledgements}


\begin{thebibliography}{}
\bibitem[2006]{ALDERING06}
Aldering G., Antilogus P., Bailey S. et al., 2006, ApJ, 650, 510
\bibitem[1994]{AF94}
Alexander, D. R., Ferguson J. W., 1994, ApJ, 437, 879
\bibitem[1982]{ARN82}
Arnett, W.D., 1982, ApJ, 253, 785
\bibitem[2003]{BERGER03}
Berger, E., Soderberg, A. M., \& Frail, D. A., 2003, IAUC, 8157, 2
\bibitem[2009]{BLONDIN09}
Blondin, S., Prieto, J.L., Patat, F.,Challis, P., Hicken, M.,
Kirshner, R.P., Matheson, T., Modjaz, M., 2009, ApJ, 693, 207
\bibitem[2011]{BIANCO11}
Bianco, F. B., Howell, D. A., Sullivan, M. et al., 2011, ApJ, 741,
20
\bibitem[1995]{BOFFI95}
Boffi, F. R. \& Branch, D., 1995, PASP, 107, 347
\bibitem[1988]{BOFFIN88}
Boffin, H. M. J. \& Jorissen, A., 1988, A\&A, 205, 155
\bibitem[1995]{BRANCH95}
Branch, D., Livio, M., Yungelson L.R. et al., 1995, PASP, 107,
1019
\bibitem[2004]{BRA04}
Branch, D., 2004, Nature, 431, 1044
\bibitem[2012a]{BROWN12a}
Brown P.J., Dawson K.S., de Pasquale M. et al., 2012a, ApJ, 753,
22
\bibitem[2012b]{BROWN12b}
Brown P.J., Dawson K.S., Harris D.W. et al., 2012b, 749, 18
\bibitem[2014]{BROWN14}
Brown, P.J., 2014, ApJL, 796, L18
\bibitem[2013]{BURKEY13}
Burkey, M.T., Reynolds, S.P., Borkowski, K.J., Blondin, J.M.,
2013, ApJ, 764, 63
\bibitem[2015]{CAOY15}
Cao, Y., Kulkarni, S.R., Howell, D.A. et al., 2015, Nature, 521,
328
\bibitem[1982a]{CHEVALIER82a}
Chevalier, R.A., 1982a, ApJ, 258, 790
\bibitem[1982b]{CHEVALIER82b}
Chevalier, R.A., 1982b, ApJ, 259, 302
\bibitem[1990]{CHEVALIER90}
Chevalier, R.A., 1990, in Supernovae, ed. A.G. Petschek (New York,
NY: Springer-Verlag), 91
\bibitem[2003]{CHEVALIER03}
Chevalier, R. A., \& Fransson, C. 2003, in Supernovae and
Gamma-Ray Bursters, ed. K. Weiler (New York, NY: Springer-Verlag),
171
\bibitem[2006]{CHEVALIER06}
Chevalier, R.A. \& Fransson, C., 2006, ApJ, 651, 381
\bibitem[2006]{CHEVALIER06b}
Chevalier R.A., Fransson C. Nymark T.K., 2006, ApJ, 641, 1029
\bibitem[2007]{CHE07}
Chen, X., Tout, C.A., 2007, ChJAA, 7, 2, 245
\bibitem[2011]{CHENXF11}
Chen, X., Han, Z., Tout, C.A., 2011, ApJ, 735, L31
\bibitem[2012]{CHENXF12}
Chen, X., Jeffery, C.S., Zhang, X., Han, Z., 2012, ApJL, 775, l9
\bibitem[2007]{CHENWC07}
Chen, W., Li, X., 2007, ApJ, 658, L51
\bibitem[2009]{CHENWC09}
Chen, W., Li, X., 2009, ApJ, 702, 686
\bibitem[2012]{CHIOTELLIS12}
Chiotellis, A., Schure, K. M., Vink, J., 2012, A\&A, 537, A139
\bibitem[2012]{CHOMIUK12}
Chomiuk, L., Soderberg, A., Moe, M. et al., 2012, ApJ, 750, 164
\bibitem[1986]{CHUGAI86}
Chugai, N.N., 1986, SvA, 30, 563
\bibitem[2004a]{CHUGAI04}
Chugai N.N., Yungelson L.R., 2004a, Astronomy Letters, 30, 65
\bibitem[2004b]{CHUGAI04b}
Chugai, N.N., Chevalier, R.A., Lundqvist, P., 2004, MNRAS, 355,
627
\bibitem[2014]{DIMITRIADIS14}
Dimitriadis, G., Chiotellis, A., Vink, J., 2014, MNRAS, 443, 1370
\bibitem[2011]{DISTEFANO11}
Di Stefano, R., Voss, R., Claeys, J. S. W., 2011, ApJ, 738, L1
\bibitem[1995]{ECK95}
Eck, C. R., Cowan, J. J., Roberts, D. A. et al., 1995, ApJ, 451,
L53
\bibitem[1971]{EGG71}
Eggleton, P.P., 1971, MNRAS, 151, 351
\bibitem[1972]{EGG72}
Eggleton, P.P., 1972, MNRAS, 156, 361
\bibitem[1973]{EGG73}
Eggleton, P.P., 1973, MNRAS, 163, 279
\bibitem[2012a]{FOLEY12}
Foley, R. J., Simon, J. D., Burns, C. R. et al., 2012a, ApJ, 752,
101
\bibitem[2012b]{FOLEY12b}
Foley, R.J., Challis, P.J., Filippenko, A.V. et al., 2012b, ApJ,
744, 38
\bibitem[1996]{FRANSSON96}
Fransson, C., Lundqvist, P., Chevalier, R.A., 1996, ApJ, 461, 993
\bibitem[2011]{GANESHALINGAM11}
Ganeshalingam, M., Li, W., Filippenko, A. V., 2011, MNRAS, 416,
2607
\bibitem[2001]{GOLGHABER01}
Goldhaber, G., Groom, D.E., Kim, A. et al., 2001, ApJ, 558, 359
\bibitem[2015]{GRAHAM15}
Graham, M.L., Valenti, S.;, Fulton, B.J. et al., 2015, ApJ, 801,
136
\bibitem[1996]{HAC96}
Hachisu, I., Kato, M., Nomoto, K., ApJ, 1996, 470, L97
\bibitem[1999a]{HAC99a}
Hachisu, I., Kato, M., Nomoto, K., Umeda, H., 1999a, ApJ, 519, 314
\bibitem[1999b]{HAC99b}
Hachisu, I., Kato, M., Nomoto, K., 1999b, ApJ, 522, 487
\bibitem[2001]{HK01}
Hachisu, I., Kato, M., 2001, ApJ, 558, 323
\bibitem[2003a]{HK03a}
Hachisu, I., Kato, M., 2003a, ApJ, 588, 1003
\bibitem[2003b]{HK03b}
Hachisu, I., Kato, M., 2003b, ApJ, 590, 445
\bibitem[2008a]{HKN08}
Hachisu, I., Kato, M., Nomoto, K., 2008a, ApJ, 679, 1390
\bibitem[2008b]{HKN08b}
Hachisu, I., Kato, M., Nomoto, K., 2008b, ApJ, 683, L127
\bibitem[2012]{HKSN12}
Hachisu, I., Kato, M., Saio, H., Nomoto, K., 2012, ApJ, 744, 69
\bibitem[2003]{HAM03}
Hamuy, M. et al., 2003, Nature, 424, 651
\bibitem[1994]{HAN94}
Han, Z., Podsiadlowski, P., Eggleton, P.P., 1994, MNRAS, 270, 121
\bibitem[1998]{HAN98}
Han, Z., 1998, MNRAS, 296, 1019
\bibitem[2000]{HAN00}
Han, Z., Tout, C.A., Eggleton, P.P., 2000, MNRAS, 319, 215
\bibitem[2004]{HAN04}
Han, Z., Podsiadlowski, Ph., 2004, MNRAS, 350, 1301
\bibitem[2006]{HAN06}
Han, Z., Podsiadlowski, Ph., 2006, MNRAS, 368, 1095
\bibitem[2008]{HAN08}
Han, Z., 2008, ApJ, 677, L109
\bibitem[2011]{HANCOCK11}
Hancock, P. P., Gaensler, B. M., Murphy, T., 2011, ApJ, 735, L35
\bibitem[2010]{HAYDEN10}
Hayden, B. T., Garnavich, P. M., Kasen, D. et al., 2010, ApJ, 722,
1691
\bibitem[2000]{HN00}
Hillebrandt, W., Niemeyer, J.C., 2000, ARA\&A, 38, 191
\bibitem[2009]{HOWEL09}
Howell, D.A. et al., 2009, arXiv: 0903.1086
\bibitem[2007]{HUGHES07}
Hughes, J. P., Chugai, N., Chevalier, R. et al., 2007, ApJ, 670,
1260
\bibitem[1960]{HF60}
Hoyle, F. \& Fowler, W.A., 1960, ApJ, 132, 565
\bibitem[1984]{IT84}
Iben, I., Tutukov, A.V., 1984, ApJS, 54, 335
\bibitem[1996]{IR96}
Iglesias, C. A., Roger,s F. J., 1996, ApJ, 464, 943
\bibitem[2012]{ILKOV12}
Ilkov, M., Soker, N., 2012, MNRAS, 419, 1695
\bibitem[2006]{IMMLER06}
Immler, S., Brown, P.J., Milne, P. et al., 2006, ApJ, 648, L119
\bibitem[2007]{IMMLER07}
Immler, S., Brown, P.J., Milne, P. et al., 2007, ApJ, 664, 435
\bibitem[2011]{JUSTHAM11}
Justham, S., 2011, ApJ, 730, L34
\bibitem[2010]{KASEN10}
Kasen, D., 2010, ApJ, 708, 1025
\bibitem[2011]{KASHI11}
Kashi, A., Soker, N., 2011, MNRAS, 417, 1466
\bibitem[2004]{KH04}
Kato, M., Hachisu I., 2004, ApJ, 613, L129
\bibitem[2004]{KOTAK04}
Kotak R., Meikle W.P.S., Adamson S. et al., 2004, MNRAS, 354, L13
\bibitem[2015]{KUTSUNA15}
Kutsuna, M. \& Shigeyama, T., 2015, PASJ, \textbf{67, 5}4, arXiv:
1504.1234
\bibitem[2013]{KUSHNIR13}
Kushnir, D., Katz, B., Dong, S. et al., 2013, ApJ, 778, L37
\bibitem[2000]{LAN00}
Langer, N., Deutschmann, A., Wellstein, S. et al., 2000, A\&A,
362, 1046
\bibitem[2000]{LEI00}
Leibundgut, B., 2000, A\&ARv, 10, 179
\bibitem[2007]{LEONARDO07}
Leonard D.C., 2007, ApJ, 670, 1275
\bibitem[1997]{LI97}
Li, X.D., van den Heuvel, E.P.J., 1997, A\&A, 322, L9
\bibitem[2015]{LIUZW15}
Liu, Z., Moriya T.J., Stancliffe, R.J., Wang, B., 2015, A\&A, 574,
A12
\bibitem[1999]{LIVIO99}
Livio, M., in Truran, J., Niemeyer, T., eds, Type Ia Suppernova:
Theory and Cosmology. Cambridge Univ. Press , New York, 1999, p.33
\bibitem[1995]{LIVIN95}
Livne, E. \& Arnett, D., 1995, ApJ, 452, 62
\bibitem[2009]{LGL09}
L\"{u}, G., Zhu, C. Wang, Z., Wang, N., 2009, MNRAS, 396, 1086
\bibitem[1988]{LUNDQVIST88}
Lundqvist, P. \& Fransson, C., 1988, A\&A, 192, 221
\bibitem[2013]{LUNDQVIST13}
Lundqvist, P., Mattila, S., Sollerman, J. et al., 2013, MNRAS,
435, 329
\bibitem[2015]{LUNDQVIST15}
Lundqvist, P., Nyholm, A., Taddia, F. et al., 2015, arXiv:
1502.0589
\bibitem[2016]{MAEDA16}
Maeda K., Tajitsu A., Kawabata K.S. et al., 2016, ApJ, 816, 57
\bibitem[2014]{MAOZ14}
Maoz D., Mannucci F., Nelemans G., 2014, ARA\&A, 52, 107
\bibitem[2000]{MAR00}
Marietta, E., Burrows, A., Fryxell, B., 2000, ApJS, 128, 615
\bibitem[2015]{MARION15}
Marion, G.H., Brown, P.J., Vink\'{o}, J. et al., 2015, arXiv:
1507.7261
\bibitem[2012]{MARGUTTI12}
Margutti, R., Soderberg, A. M., Chomiuk, L. et al., 2012 ApJ, 751,
134
\bibitem[2014]{MARGUTTI14}
Margutti, R., Parrent, J., Kamble, A. et al., 2014, ApJ, 790, 52
\bibitem[2005]{MATTILA05}
Mattila, S., Lundqvist, P., Sollerman, J., et al. 2005, A\&A, 443,
649
\bibitem[2013]{MAGUIRE13}
Maguire, K., Sullivan, M., Patat, F. et al., 2013, MNRAS, 436, 222
\bibitem[2007]{MENGXC07}
Meng, X., Chen, X., Han, Z., 2007, PASJ, 59, 835
\bibitem[2008]{MENG08}
Meng, X., Chen, X., Han, Z., 2008, A\&A, 487, 625
\bibitem[2009]{MENG09}
Meng, X., Chen, X., Han, Z., 2009, MNRAS, 395, 2103
\bibitem[2010a]{MENGYANG10}
Meng, X., Yang, W., 2010a, ApJ, 710, 1310
\bibitem[2010b]{MENGYANG10c}
Meng, X., Yang, W., 2010b, MNRAS, 401, 1118
\bibitem[2011]{MENGYANG11}
Meng, X., Yang, W., 2011, RAA, 11, 965
\bibitem[2013]{MENGPOD13}
Meng, X., Podsiadlowski, Ph., 2013, ApJL, 778, L35
\bibitem[2015]{MENG15}
Meng, X., Gao, Y., Han, Z., 2015, IJMPD, 24, 14, 1530029
\bibitem[2010]{MIKOLAJEWSKA10}
Mikolajewska, J., 2010, arXiv: 1011.5657
\bibitem[2012]{MIKOLAJEWSKA11}
Mikolajewska, J., 2012, BaltA, 21, 5
\bibitem[2013]{MOHAMED13}
Mohamed, S., Booth, R., Podsiadlowski, P., 2013, in Di Stefano R.,
Orio M., MoeM., eds, Proc. IAU Symp. 281, Binary Paths to Type Ia
Supernovae Explosions. Cambridge Univ. Press, Cambridge, p. 195
\bibitem[1984]{NTY84}
Nomoto, K., Thielemann, F-K., Yokoi, K., 1984, ApJ, 286, 644
\bibitem[1999]{NOM99}
Nomoto, K., Umeda, H., Hachisu, I. Kato, M., Kobayashi, C.,
Tsujimoto, T., 1999, in Truran J., Niemeyer T., eds, Type Ia
Suppernova :Theory and Cosmology.Cambridge Univ. Press, New York,
p.63
\bibitem[2003]{NOM03}
Nomoto, K., Uenishi, T., Kobayashi, C. Umeda, H., Ohkubo, T.,
Hachisu, I., Kato, M., 2003, in Hillebrandt W., Leibundgut B.,
eds, From Twilight to Highlight: The Physics of supernova,
ESO/Springer serious ``ESO Astrophysics Symposia'' Berlin:
Springer, p.115
\bibitem[2015]{OLLING15}
Olling, R.P., Mushotzky, R., Shaya, E.J. et al., 2015, Nature,
521, 332
\bibitem[2012]{PAN12}
Pan, K., Ricker, P., Taam, R., 2012, ApJ, 750, 151
\bibitem[2006]{PANAGIA06}
Panagia, N., Van Dyk, S. D., Weiler, K. W. et al., 2006, ApJ, 646,
369
\bibitem[2007]{PAT07}
Patat, F. et al., 2007, Science, 317, 924
\bibitem[2011]{PATAT11}
Patat, F., Chugai, N. N., Podsiadlowski, Ph. et al., 2011, A\&A,
530, A63
\bibitem[2013]{PATAT13}
Patat F., Cordiner M.A., Cox N.L.J. et al., 2013, A\&A, 549, A62
\bibitem[2012]{PATNAUDE12}
Patnaude, D.J., Badenes, C., Park, S., Laming, J. M., 2012, ApJ,
756, 6
\bibitem[2008]{PODSIADLOWSKI08}
Podsiadlowski, P., 2008, ASP Conf. Ser., 401, 63
\bibitem[1999]{PER99}
Perlmutter, S. et al., 1999, ApJ, 517, 565
\bibitem[2014]{PEREZ14}
P\'{e}rez-Torres, M., Lundqvist, P., Beswick, R. et al., 2014,
ApJ, 792, 38
\bibitem[1995]{POL95}
Pols, O.R., Tout C.A., Eggleton P.P. et al., 1995, MNRAS, 274, 964
\bibitem[1997]{POL97}
Pols, O.R., Tout C.A., Schr\"{o}der K.P. et al., 1997, MNRAS, 289,
869
\bibitem[1998]{POL98}
Pols, O.R., Schr\"{o}der K.P., Hurly J.R. et al., 1998, MNRAS,
298, 525
\bibitem[2009]{RASKIN09}
Raskin, C., Timmes, F.X., Scannapieco, E. et al., 2009, MNRAS,
399, L156
\bibitem[1975]{REIMERS75}
Reimers, D., 1975, Mem. R. Soc. li\`{e}ge, 6i\`{e}me Serie, 8, 369
\bibitem[1998]{RIE98}
Riess, A. et al., 1998, AJ, 116, 1009
\bibitem[2009]{RUITER09}
Ruiter, A.J., Belczynski, K., Fryer, C., 2009, ApJ, 699, 2026
\bibitem[2011]{RUITER11}
Ruiter, A.J., Belczynski, K., Sim, S.A. et al., 2011, MNRAS, 417,
408
\bibitem[2014]{RUIZLAPUENTE14}
Ruiz-Lapuente, P., 2014, NewAR, 62, 15
\bibitem[2012]{RUSSELL12}
Russell, B. R. \& Immler, S., 2012, ApJ, 748, L29
\bibitem[1997]{SCH97}
Schr\"{o}der, K.P., Pols, O.R., Eggleton, P.P., 1997, MNRAS, 285,
696
\bibitem[1998]{SCHMIDT98}
Schmidt, B. P., Suntzeff, N. B., Phillips, M. M. et a., 1998, ApJ,
507, 46
\bibitem[2013]{SHEN13}
Shen, K.J., Guillochon, J., Foley, R.J., 2013, ApJL, 770, L35
\bibitem[2009]{SIMON09}
Simon, J. D., Gal-Yam ,A., Gnat, O. et al. 2009, ApJ, 702, 1157
\bibitem[2005]{SODERBERG05}
Soderberg, A. M., \& Frail, D. A. 2005, ATel, 663, 1
\bibitem[2006]{SOKOLOSKI06a}
Sokoloski, J. L., Luna, G. J. M., Mukai, K. et al., 2006, Nature,
442, 276
\bibitem[2015]{SOKER15}
Soker, N. 2015, MNRAS, 450, 1333, arXiv: 1501.7729
\bibitem[2011]{STERNBERG11}
Sternberg, A., Gal-Yam, A., Simon, J.D. et al., 2011, Science,
333, 856
\bibitem[2014]{STERNBERG14}
Sternberg, A., Gal-Yam, A., Simon, J.D. et al., 2014, MNRAS, 443,
1849
\bibitem[2003]{STOCKDALE03}
Stockdale, C. J., Sramek, R. A., Weiler, K. W., et al. 2003, IAUC,
8157, 3
\bibitem[1998]{SUTHERLAND98}
Sutherland, R.S., 1998, MNRAS, 300, 321
\bibitem[1988]{TOUT88}
Tout, C. A., \& Eggleton, P. P. 1988, ApJ, 334, 357
\bibitem[1991]{TOUT91}
Tout, C. A. \& Hall, D. S., 1991, MNRAS, 253, 9
\bibitem[2008]{TOTANI08}
Totani, T., Morokuma, T., Oda, T. et al., 2008, PASJ, 60, 1327
\bibitem[2015]{TSEBRENKO15}
Tsebrenko, D., Soker, N., 2015, MNRAS, 447, 2568
\bibitem[2011]{TUCKER11}
Tucker, B. E., 2011, ApSS, 335, 223
\bibitem[2004]{UENISHI04}
Uenishi T., Suzuki T., Nomoto K. et al., 2004, Rev. Mex. A\&A, 20,
219
\bibitem[1999]{UME99}
Umeda, H., Nomoto, K., Yamaoka, H. et al., 1999, ApJ, 513, 861
\bibitem[2004]{WANGL04}
Wang L., Baade D., H\"{o}flich P. et al., 2004, ApJ, 604, L53
\bibitem[2009a]{WANGB09a}
Wang, B., Meng, X., Chen, X., Han, Z., 2009a, MNRAS, 395, 847
\bibitem[2009b]{WANGB09b}
Wang, B., Chen, X., Meng, X., Han, Z., 2009b, ApJ, 701, 1540
\bibitem[2010]{WANGB10}
Wang, B., Li, X., Han, Z., 2010, MNRAS, 401, 2729
\bibitem[2012]{WANGB12}
Wang, B. \& Han, Z., 2012, NewAR, 56, 122
\bibitem[1984]{WEB84}
Webbink, R.F., 1984, ApJ, 277, 355
\bibitem[1986]{WEILER86}
Weiler, K.W., Sramek, R.A., Panagia, N. et al., 1986, ApJ, 301,
790
\bibitem[1973]{WI73}
Whelan, J., Iben, I., 1973, ApJ, 186, 1007
\bibitem[2014]{WILLIAMS14}
Williams, B.J., Borkowski, K.J., Reynolds, S.P. et al., 2014, ApJ,
790, 139
\bibitem[1994]{WOOSLEY94}
Woosley, S.E., Weaver, T.A., 1994, ApJ, 423, 371
\bibitem[2004]{YOON04}
Yoon, S.-C. \& Langer, N., 2004, A\&A, 419, 623
\bibitem[2005]{YOON05}
Yoon, S.-C. \& Langer, N., 2005, A\&A, 435, 967
\bibitem[1995]{YUN95}
Yungelson, L., Livio, M., Tutukou, A. Kenyon, S.J., 1995, ApJ,
447, 656

\end{thebibliography}
\end{document}